\documentclass[%
 reprint,
superscriptaddress,
 amsmath,amssymb,
 aps,
prb,
]{revtex4-2}

\usepackage{graphicx}
\usepackage{physics}
\usepackage{dcolumn}
\usepackage{bm}
\usepackage{tabularx}
\usepackage{xcolor} 
\usepackage[colorlinks,linkcolor=blue,citecolor=blue,urlcolor=blue]{hyperref}
\usepackage{natbib}

\begin{document}

\preprint{APS/123-QED}

\title{High-fidelity iSWAP gate with Double Transmon Coupler}

\author{Tarush Tiwari}
\affiliation{Department of Physics and Applied Physics, University of Massachusetts, Lowell, MA, 01854, USA}
 \affiliation{Quantum, Photonics and Computing, RTX BBN Technologies, Cambridge, MA, 02138, USA}
\author{Sudhir K. Sahu} 
\altaffiliation{Present Address: Department of Physics, University of Colorado, Boulder, Colorado 80309, USA}
\affiliation{Department of Physics and Applied Physics, University of Massachusetts, Lowell, MA, 01854, USA}
\affiliation{National Institute of Standards and Technology, Boulder, CO 80305, USA}
\author{Guilhem Ribeill}%
\affiliation{%
Quantum, Photonics and Computing, RTX BBN Technologies, Cambridge, MA, 02138, USA}%
\author{Michael Senatore}
\affiliation{Air Force Research Laboratory, Information Directorate, Rome, New York 13441, USA}
\author{Matthew D. LaHaye}
\affiliation{Air Force Research Laboratory, Information Directorate, Rome, New York 13441, USA}
\author{Raymond W. Simmonds} 
\affiliation{National Institute of Standards and Technology, Boulder, CO 80305, USA}
\author{Daniel L. Campbell}
\email{daniel.campbell.22@us.af.mil}
\affiliation{Air Force Research Laboratory, Information Directorate, Rome, New York 13441, USA}
\author{Archana Kamal}
\email{archana.kamal@northwestern.edu}
\affiliation{Department of Physics and Applied Physics, University of Massachusetts, Lowell, MA, 01854, USA}
\affiliation{Department of Physics and Astronomy, Northwestern University, Evanston, IL 60208, USA}
\author{Leonardo Ranzani}%
 \email{leonardo.ranzani@rtx.com}
\affiliation{%
Quantum, Photonics and Computing, RTX BBN Technologies, Cambridge, MA, 02138, USA}%

\date{\today}

\begin{abstract}
Entangling operations are at the heart of all approaches to quantum information processing. Parametric gates, in particular, offer a versatile solution to strongly couple off-resonant superconducting qubits with suppressed parasitic crosstalk to spectator qubits due to frequency-selective activation. In this work, we demonstrate a parametric iSWAP gate between two transmon qubits using the recently developed double transmon coupler (DTC). The DTC supports robust internally-defined cancellation point (``off'' state) for static interactions, while simultaneously mediating a fast parametric coupling between data qubits that can be deployed for high-fidelity two-qubit operations. We use robust phase estimation to calibrate non-commuting error terms in the parametric iSWAP gate, and achieve a 99.827\% gate fidelity in 40\,ns without any numerical optimization. The circuit architecture and calibration techniques developed here are extensible to other gate implementations and qubit modalities, paving the way towards resource-efficient quantum information processing.
\end{abstract}

\maketitle
%
\section{\label{sec:level1}Introduction}
%
Error-corrected quantum computing platforms rely on high-fidelity gates to exponentially suppress errors during computation. This continues to motivate several advances in coupling architectures and quantum control techniques with the central aim of improving single- and two-qubit gate fidelities. In the context of superconducting qubits, two-qubit gates are typically implemented using either static or flux-tunable couplings, with the latter option typically enabling faster gates and high on/off ratios~\citep{wu2024modular,valles2025optimizing,li2024realization,kim2026tunable,koshino2025galvanically,xu2026tunable}. In addition, by performing a time-dependent modulation of the coupler flux at a linear combination of mode frequencies, strong parametric interactions can be implemented between data qubits~\citep{zhang2024tunable,jin2025superconducting,chakraborty2025tunable,subramanian2023efficient} without tuning the respective qubit energy levels in resonance. This strategy for multi-qubit control is especially advantageous when scaling to a large number of qubits, since it reduces the risk of frequency collisions during gate operations. In addition, parametric gates also reduce frequency crowding, increase qubit connectivity and enable effective three-qubit or multi-qubit gates~\citep{warren2023extensive,subramanian2023efficient}. 

Despite these numerous advantages, while resonant gates using tunable couplers with fidelities up to 99.9\% have been demonstrated~\citep{sung2021realization}, achieving similar performance with parametric gates is harder as they involve more complicated drive optimization and lengthy tuneup procedures that result in lower average fidelities. This is especially true for high-fidelity parametric iSWAPs since exchange-type interactions do not commute with single-qubit evolution (Z rotations), placing extra stringent demands on phase stability and cross-Kerr (or ZZ) compensation. In view of recent works, which have highlighted the unique advantages offered by addition of fast iSWAP gates to quantum control toolkit, this situation has turned particularly vexing! For example, iSWAP-based two-qubit operations can reduce the circuit depth for analog quantum optimization algorithms \cite{Abrams2020}, ease resource requirements for quantum chemistry simulations~\citep{Ganzhorn2019}, and enable surface code implementations with reduced coupler count per physical qubit~\citep{mcewen2023relaxing}. The iSWAP gate is also more robust against leakage errors, as compared to CNOT or CZ, since its operation involves only computational states. 

In this paper, we demonstrate parametric iSWAP between two qubits with 99.827\% fidelity in $40\,$ns by employing a double-transmon coupler~\citep{campbell2023modular,goto2022double,li2024realization,campbell2026transmon,heunisch2023tunable,li2025capacitively,yuan2026tunable} to achieve fast swap rates while suppressing static and dynamic cross-Kerr interactions during the gate operation. In addition, we develop a robust phase estimation routine to accurately and efficiently calibrate the iSWAP gate by selectively amplifying arbitrary gate errors. Our calibration routine sequentially measures and corrects unknown gate parameters, precluding the need for any numerical pulse optimization or repeated randomized benchmarking. Phase estimation has previously been used to calibrate single-qubit gates~\citep{kimmel2015robust} and, very recently, two-qubit CZ gates~\citep{rudinger2025heisenberg}. However, unlike previous works, error terms such as single-qubit Stark shifts do not commute with the exchange interaction and, therefore, cannot be straightforwardly amplified by repeating the iSWAP operation. In fact, these errors affect both the eigenvalues and eigenvectors of the unitary representing the gate and repeated application of the unknown gate causes the state vector to rotate, suppressing the desired error amplification. Here we solve this problem by designing suitable compound gates that selectively amplify specific error terms in the iSWAP, while canceling unwanted state vector rotations. 

The paper is organized as follows: Sec.~\ref{sec:level2} details the device design and basic operating principle of double-transmon coupler (DTC). Sec.~\ref{sec:level3} includes the characterization of static coupling cancellation and crosstalk suppression enabled by the DTC . This is followed by a detailed discussion on calibration and characterization of the parametric iSWAP gate in Sec.~\ref{sec:level4}. Sec.~\ref{sec:level5} concludes the main text with a brief summary of the key results. Additional theoretical and experimental details are included in appendices \ref{sec:appendix_setup}-\ref{sec:appendix_RBCliffords}.
%
\section{\label{sec:level2}Device description}
%
\begin{figure}[t!]
\begin{centering}
    \includegraphics[width=0.96\linewidth]{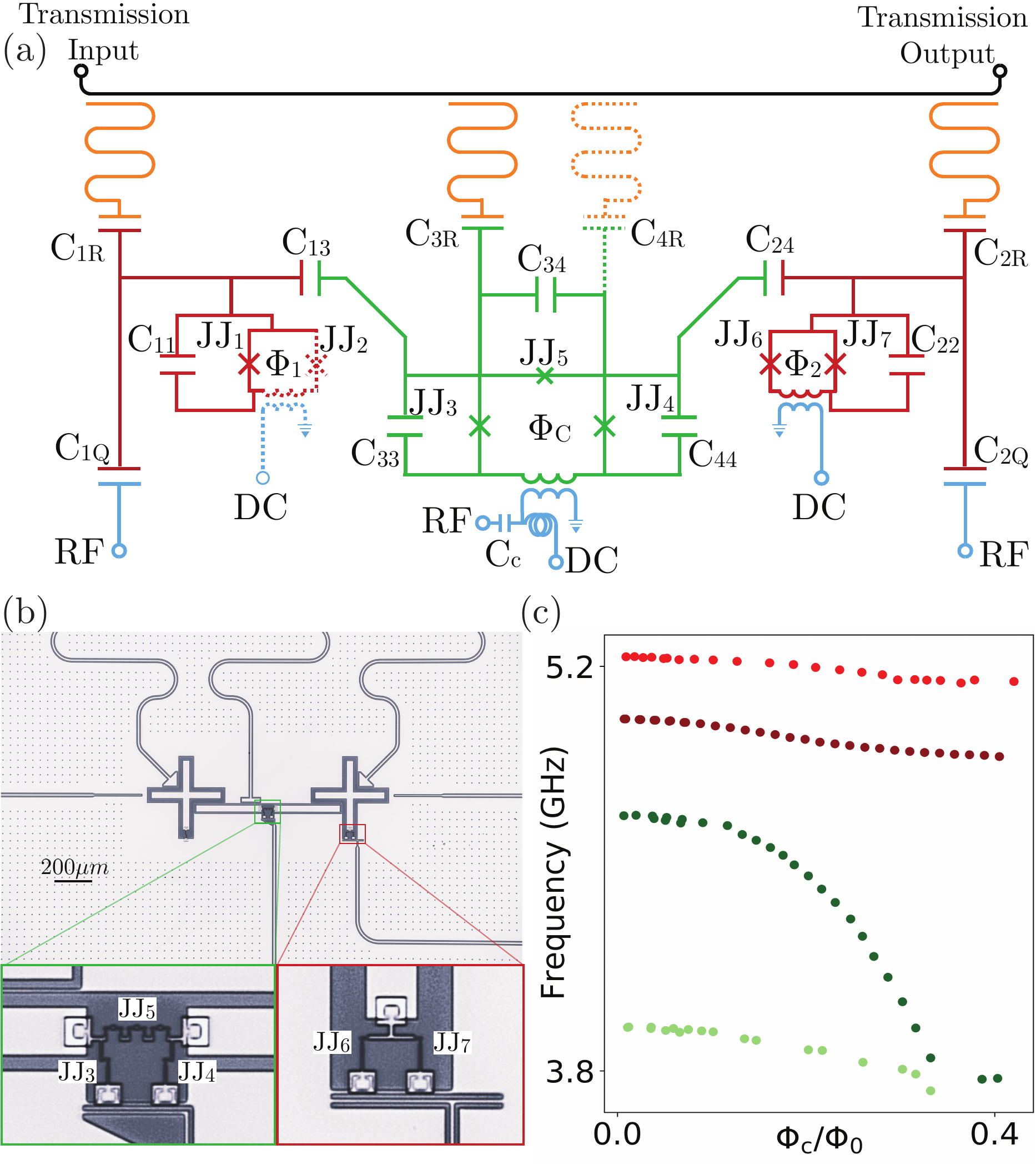}
    \caption{\label{Fig:Fig1}(a) Circuit schematic of the device, consisting of two data qubits ($Q_1$,$Q_2$, red) connected by a double transmon coupler ($Q_3$,$Q_4$, green). The circuit elements shown in dashed lines are absent in the picture of the specific sample shown in (b), though were present on the measured device.  (b) Optical micrographs of a representative device. Insets show the junction layout of the double-transmon coupler a.k.a. DTC (left) and data qubit (right). (c) Two-tone spectroscopy of data (dark and bright red) and coupler (dark and bright green) transmons measured as a function of coupler flux bias $\Phi_{c}$.}
\end{centering}
\end{figure}

The device layout, shown in Fig.~\ref{Fig:Fig1}(a), consists of two tunable transmon data qubits $Q_1$ and $Q_2$ and a double-transmon coupler comprising two additional transmon qubits $Q_3$ and $Q_4$. Each data qubit has dedicated dc flux bias lines inductively coupled to Josephson junction loops formed by $JJ_1$ and $JJ_2$ for $Q1$ and by $JJ_6$ and $JJ_7$ for $Q2$, and microwave controls lines coupled via $C_{1Q}$ and $C_{2Q}$ to implement single qubit rotations. Each of the four transmon qubits is capacitively coupled to a dedicated readout resonator to enable single-shot readout. 

The coupler transmons are connected in a three-junction loop ($JJ_3$, $JJ_4$, $JJ_5$), which is flux biased using a separate dc control line. An RF control line is combined with the coupler flux bias to facilitate parametric flux modulation, which is utilized to implement  the parametric iSWAP gate presented in this work (see appendix~\ref{sec:appendix_setup}). Junctions $JJ_3$ and $JJ_4$ have nominally identical critical currents while the coupling junction $JJ_5$ is designed to have one third of the critical current of junctions $JJ_{3,4}$. The two transmon qubits that form the coupler, $Q_3$ and $Q_4$, are strongly coupled via a Josephson junction ($JJ_5$) whose participation is set by the coupler flux bias $\Phi_{c}$. This results in a flux-dependent hybridization of the two coupler transmons which, in turn, determines the net interaction between them through a competition between the capacitive $g_C$ and inductive $g_L(\Phi_c)$ interactions,
\begin{subequations}
\begin{align}
g_L (\Phi_c)&=\frac{4E_{J5}\cos(2\pi\Phi_c/\Phi_0)\sqrt{E_{c3}E_{c4}}}{\hbar^2\sqrt{\omega_{3}\omega_{4}}}, \\
g_C&=\frac{C_{34}\sqrt{\omega_{3}\omega_{4}}}{2\sqrt{(C_{33}+C_{34})(C_{44}+C_{34})}}.
\end{align}
\label{eq:couplings}%
\end{subequations}
Here $E_{J5}$ is the Josephson energy of $JJ_5$ and $E_{c3,4}$, $\omega_{3,4}$ denote the anharmonicities and frequencies of the coupler qubits respectively. The capacitive and inductive interactions have opposite signs and, therefore, the net interaction between coupler qubits is given by $g_C - g_L(\Phi_c)$. The full Hamiltonian for the DTC can then be written as~\citep{campbell2023modular},
\begin{align}
    &H/\hbar = \sum_{j=1,2,3,4} \left( \omega_{j} q^{\dagger}_{j} q_{j} + \frac{\alpha_j}{2}q^{\dagger}_{j}q^{\dagger}_{j}q_{j}q_{j}\right)      \nonumber\\  & \qquad \qquad+ (g_C - g_L (\Phi_c))(q^{\dagger}_{3}q_{4} + q_{3}q^{\dagger}_{4}) 
      \nonumber\\  & \qquad \qquad + g_{13}(q^{\dagger}_{1}q_{3} + q_{1}q^{\dagger}_{3} ) +  g_{24}(q^{\dagger}_{2}q_{4} + q_{2}q^{\dagger}_{4} ),
     \label{eq:corot}
\end{align}
where $q_{j}$ and $q^{\dagger}_{j}$ index the annihilation and creation operators associated with the bare transmon qubit modes. The coupler qubits are designed to nominally have the same bare frequencies ($\omega_{3}$ = $\omega_{4}$). Each data qubit is coupled to its nearest coupler qubit, with couplings $g_{13}$ and $g_{24}$ between qubit pairs $Q_1$-$Q_3$, $Q_2$-$Q_4$ set by capacitances $C_{13}$ and $C_{24}$ respectively. By tuning the inductive coupling $g_L$, we can control the normal mode frequencies of the coupler and, therefore, the effective flux-dependent coupling $g_{\rm eff} (\Phi_c)$ between the data qubits $Q_1$ and $Q_2$~\citep{campbell2023modular},
\begin{align}
g_{\textrm{eff}} (\Phi_c)  &\approx  g_{13} g_{24} (g_C-g_L(\Phi_c)) \times\nonumber\\
 & \sum_{j=3,4} \frac{1}{2D_j^2(\Phi_c)}\left[1+\frac{(g_C+g_L(\Phi_c))}{(g_C-g_L(\Phi_c))}\frac{\Delta_j}{\bar{\omega}}\right],
\label{eq:geff}
\end{align}
where $D_j^2(\Phi_c) = \Delta_j^2 - \delta^2 - (g_C-g_L(\Phi_c))^2$, with ${\Delta_j = \omega_{j} - \bar{\omega}}$ being the detuning of the coupler qubit frequency $\omega_{j}$ from the average frequency of the coupler $2\bar{\omega} = \omega_{3} + \omega_{4}$ and $2\delta = \omega_{3}-\omega_{4}$. 
An important feature of the DTC is the existence of a `cancellation flux' at which $g_{\rm eff} (\Phi_c)=0$~\citep{campbell2023modular}. The cancellation flux bias is typically close to the center of the avoided level crossing between the coupler qubits, with the exact value determined by the coupler circuit parameters. Such a decoupling point is present even for $g_C=0$, in contrast to single transmon or single SQUID couplers that typically require a finely-tuned capacitive coupling to negate the net inductive coupling and achieve cancellation~\citep{jin2025superconducting}. 

Figure~\ref{Fig:Fig1}(b) shows an optical micrograph of the device. The junctions are fabricated using double-angle aluminum evaporation on a Silicon substrate. The frequencies of data (red) and coupler (green) transmons [Fig.~\ref{Fig:Fig1}(a)] can be tuned independently using external flux lines, though we operate the data transmons at their respective sweet spots (zero flux) for gate characterization. As shown by the two-tone qubit spectroscopy in Fig.~\ref{Fig:Fig1}(c), near the cancellation flux $\Phi_C=0.33\Phi_0$ (see Sec.~\ref{sec:level2}), the data qubits have frequencies $4.91\,$GHz ($Q_1$) and $5.16\,$GHz ($Q_2$) with respective anharmonicities of $-172\,$MHz ($Q_1$) and $-164\,$MHz ($Q_2$). The coupler flux bias causes minimal detuning of the data qubits, due to weak flux dependence of data qubits mediated by capacitors $C_{13}$ and $C_{24}$. 
%
%
\section{\label{sec:level3}Static decoupling and ZZ interaction}
%
The double transmon coupler (DTC) robustly suppresses static interactions between the data qubits~\citep{goto2022double,campbell2023modular,li2024realization} at an optimal flux bias where the two hybridized coupler transmon modes induce equal and opposite couplings [see Eq.~(\ref{eq:couplings})].
Away from this cancellation flux, the DTC induces a static XX type interaction between the data qubits. However, as a consequence of the qubit anharmonicity, such XX interactions give rise to a cross-Kerr interaction~\citep{magesan2020effective,solgun2022direct}, which can be parametrized as, $H_{\rm int}=(g_{\rm zz}/4)\sigma_{\rm z1}\sigma_{\rm z2}$, by truncating the Hilbert space to computational states of two qubits. Due to this spurious ZZ interaction, each qubit frequency becomes dependent on the state of the other qubit which is particularly deleterious to single- and two-qubit gate fidelities. The static $g_{\rm zz}$ coupling strength is proportional to the effective coupling between data qubits $g_{\rm eff}$~\citep{ku2020suppression,jin2025superconducting}, as
\begin{equation}
g_{\rm zz}  (\Phi_c) = 2g_{\rm eff}^2 (\Phi_c)\frac{\alpha_1+\alpha_2}{(\Delta_{\rm RSB}-\alpha_1)(\Delta_{\rm RSB}+\alpha_2)},
\label{eq:gzz}
\end{equation}
with $\Delta_{\rm RSB} =\omega_{1}-\omega_{2}$. Since the anharmonicities $\alpha_j$ are always negative for transmon qubits, $g_{\rm eff}$ needs to be suppressed in order to achieve zero $g_{\rm zz}$. The DTC optimal flux bias where $g_{\rm eff} = 0$, naturally minimizes $g_{\rm zz}$ as well. A unique feature of the DTC is that the value of this cancellation flux is first-order insensitive to the frequencies of the data qubits, rendering DTC design modular and robust against fabrication variations~\citep{campbell2023modular}.
\begin{figure}
    \centering
    \includegraphics[width=\linewidth]{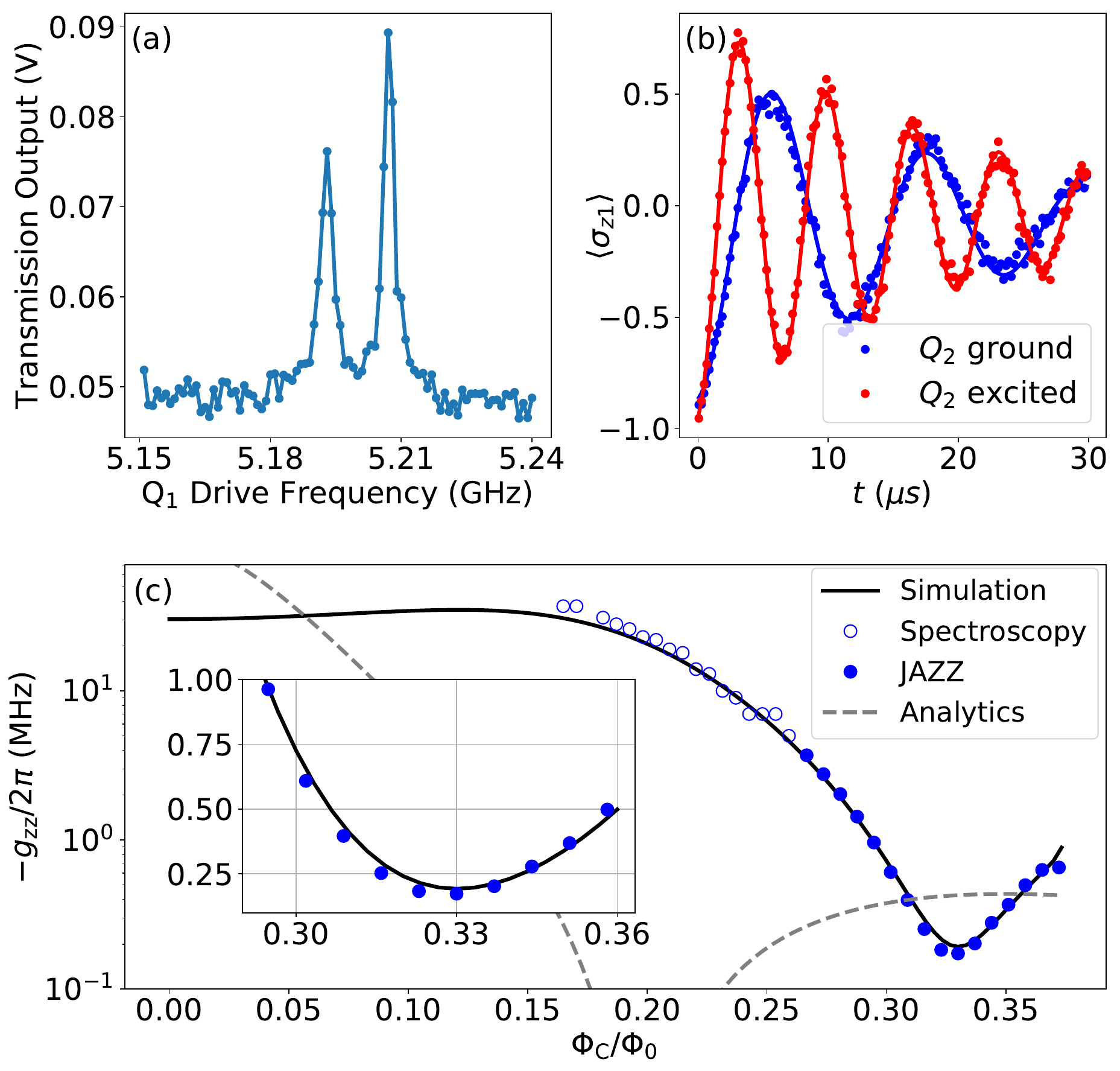}
    \caption{Static $g_{\rm zz}$ cancellation in the DTC. Measurement of the $g_{\rm zz}$ interaction using (a) simultaneous qubit spectroscopy for $g_{\rm zz}>$ 1~MHz and (b) time-domain JAZZ experiment for lower values of $g_{\rm zz}$. (c) Variation of $g_{\rm zz}$ as a function of coupler bias ($\Phi_{C}$): data (circles) vs. circuit simulation using scQubits (solid line). The error bar on the data is comparable to the size of the blue markers over the full flux range.  The simulation accurately predicts the measured $g_{\rm zz}$ with errors increasing far away from the cancellation flux. The theoretical approximation in Eq.~(\ref{eq:gzz}), which neglects the effect of higher energy levels, is also shown as for comparison (grey-dashed line). }
    \label{fig:ZZ_cancellation}
\end{figure}
Typically $g_{\rm zz}$ can be experimentally determined by measuring the dressed eigenenergies (or frequencies) of the states $|ij\rangle$ of the two data qubits as,
\begin{equation}
    \hbar g_{\rm zz} = E_{11} - E_{10} - E_{01} + E_{00}.
    \label{eq:gzzenergy}
\end{equation}
However, due to large variation of $g_{\rm zz} (\Phi_c)$ in the measured device, almost from near-zero to 30~MHz and above, we use a combination of spectroscopic [for strong $g_{\rm zz}$, Fig.~\ref{fig:ZZ_cancellation}(a)] and time-domain methods [for weak $g_{\rm zz}$, Fig.~\ref{fig:ZZ_cancellation}(b)] for its characterization. In the joint spectroscopy experiment, we sweep the frequency of the control drive on $Q_1$, while simultaneously driving $Q_2$ on resonance, which results in a double-peak feature in spectroscopy [Fig.~\ref{fig:ZZ_cancellation}(a)]. The coupling $g_{\rm zz}$ can be directly estimated from the separation between the spectroscopic peaks, as in Eq.~(\ref{eq:gzzenergy}). For lower values of $g_{\rm zz}$ (below 10~MHz), we instead use Joint Amplification of ZZ (or JAZZ) sequences~\citep{ku2020suppression,goto2022double}, where we perform two Hahn echo experiments on $Q_1$, with $Q_2$ initialized in its ground and excited states respectively, see Fig.~\ref{fig:ZZ_cancellation}. Moreover, unlike the conventional Hanh echo, we $\pi$-pulse \emph{both} qubits in the middle of the sequence. It is in the idle phase of the experiment where any cross-Kerr shift induced by the spectator qubit (the qubit not being measured) can cause the measured qubit detuning to change. By applying a $\pi$-pulse on both qubits in the middle of the sequence, unlike a conventional echo experiment where only $Q_1$ would be $\pi$-pulsed, we prevent the frequency shift due to $g_{\rm zz}$ from being echoed away. We finally extract $g_{\rm zz}$ as the difference of the Ramsey oscillation frequencies in the two experiments. In our experiment, we add a virtual Z rotation at the end of the sequence, with angle proportional to the delay between $\pi/2$ pulses to improve fitting.

We show the measured $g_{\rm zz}$ interaction as a function of coupler flux bias in Fig.~\ref{fig:ZZ_cancellation}(c). As the coupler bias approaches the cancellation flux, $\Phi_C =  0.33 \Phi_0$, we observe a minimum ${g_{\rm zz} = -2 \pi\times 220}$~kHz. We attribute the presence of a residual non-zero $g_{\rm zz}$ to higher energy levels of the coupler. We validate this using full circuit simulations performed using scQubits~\citep{groszkowski2021scqubits}, whose results show good agreement with the experimentally measured $g_{\rm zz}$ near the cancellation flux [continuous line in Fig.~\ref{fig:ZZ_cancellation}(c)]. The simulations also show that such residual cross-Kerr coupling can be further reduced by increasing the coupler frequency in order to reduce the effect of higher order modes and, in fact, can be fully eliminated by placing the coupler frequency above that of the qubits (see appendix~\ref{sec:appendix_zz}).  For the rest of the studies presented in this paper, we parked the coupler at this ``idle" point where $g_{\rm zz}$ is minimum. 

In order to assess how effective the circuit is at decoupling the data qubits at its idle point during qubit operations, we performed a simultaneous randomized benchmarking (simRB) experiment~\citep{gambetta2012characterization}. Besides verifying the impact of residual $g_{\rm zz}$ on gate fidelity, simRB is also sensitive to the presence of other nonidealities, such as drive leakage or drive-induced Stark shifts, that can occur due to simultaneous driving of the two qubits. For the simRB experiment, we initialize the qubits in their ground state and apply sequences of pairs of single-qubit gates, randomly selected from the tensor product of single-qubit Clifford groups $\mathcal{C}^{(1q)}_1 \otimes \mathcal{C}^{(1q)}_2$. We apply $(m-1)$ random gates, followed by a single recovery gate consisting of the inverse of the product of the previous gates in the sequence. We measure the expectation values of the single-qubit Z operators ($\langle \sigma_{Z1} \rangle$, $\langle \sigma_{Z2} \rangle$) as well as the two-qubit ZZ correlator ($\langle \sigma_{Z1} \sigma_{Z2} \rangle$) with increasing number of Clifford rotations ($m$). All three quantities are expected to decay exponentially and can be fitted using
\begin{equation} \label{RB_Fidelity}
    \mathcal{F} = Ap^{m} + B,
\end{equation}
where $\mathcal{F}$ is the measured observable ($\langle \sigma_{Z1} \rangle$, $\langle \sigma_{Z2} \rangle$ or $\langle \sigma_{Z1} \sigma_{Z2} \rangle$), $p$ is the depolarizing parameter, and A, B are constants. From Eq.~(\ref{RB_Fidelity}), we extract the average error rate $r=(d-1)(1-p)/d$~\citep{gambetta2012characterization}, with $d=2^n=2$ for single-qubit gates.

\begin{figure}[t!]
     \centering
     \includegraphics[width=\linewidth]{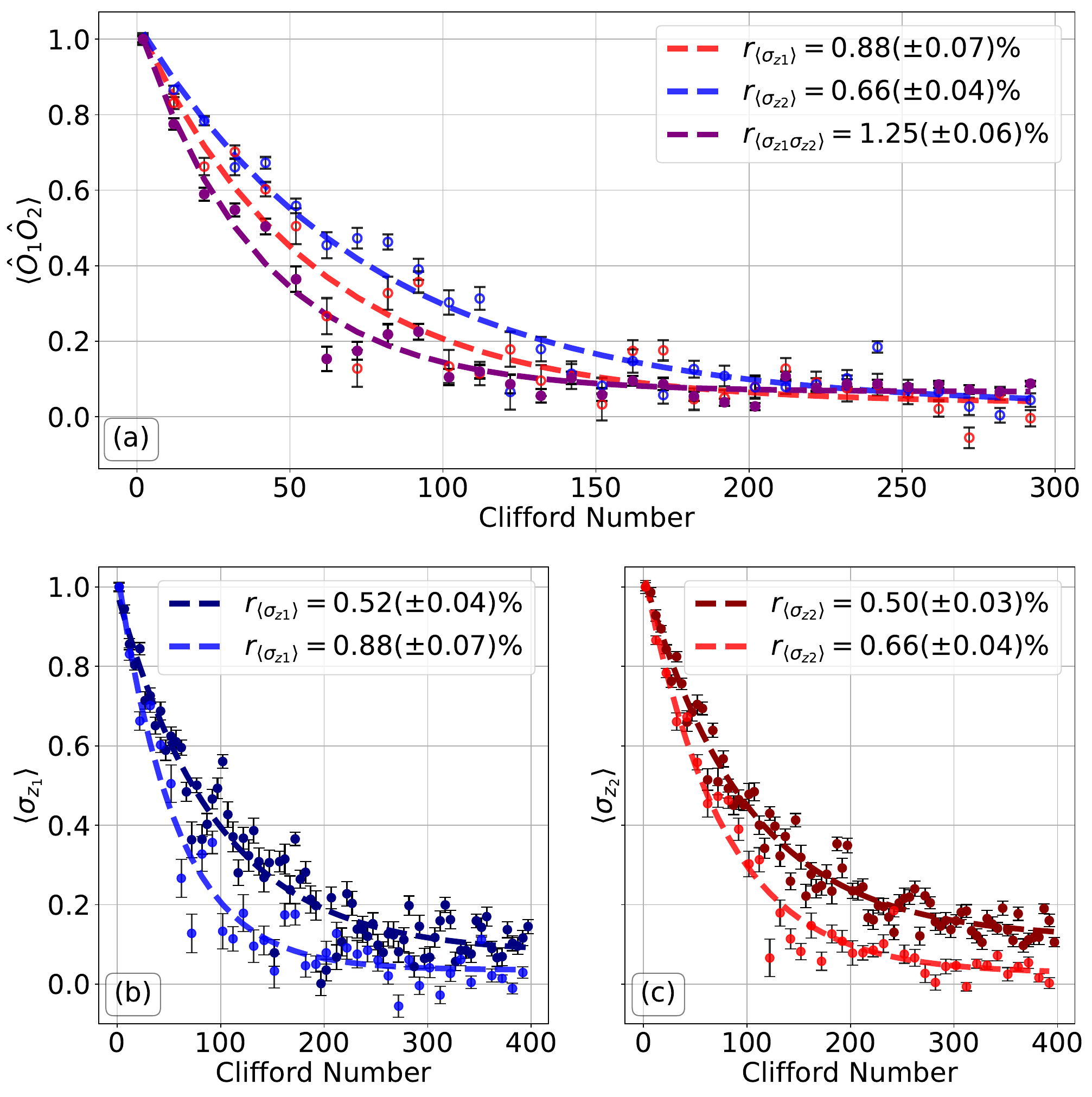}
     \caption{(a) Simultaneous two-qubit RB showing effective decoupling of the two data transmons. The product of the depolarizing parameters ($p=1-2r$) from the single-qubit measurements ($ \langle \sigma_{Z_1} \rangle$ and $ \langle \sigma_{Z2} \rangle$) lies within the error range of the correlator measurements ($\langle \sigma_{Z_1}\sigma_{Z_2} \rangle$). (b-c) Comparison of single-qubit polarizations obtained from simultaneous (dashed) and single-qubit (continuous) RBs on data qubits $Q_1$ and $Q_2$.}
     \label{fig:SingleRB}
 \end{figure}

Figure~\ref{fig:SingleRB} shows the result of the simultaneous RB study.
We observe that the product of the depolarizing parameters from single qubit $\sigma_Z$ operators is approximately equal to the depolarizing parameter of the two-qubit correlator $p_{\langle \sigma_{Z1} \rangle}p_{\langle \sigma_{Z2} \rangle} \approx p_{\langle \sigma_{Z1}\sigma_{Z2} \rangle}$ [Fig.~\ref{fig:SingleRB}(a)] confirming that we achieve effective decoupling of the two data qubits~\citep{gambetta2012characterization}.

Comparing the simRB with single-qubit RB also provides insight into `classical' crosstalk or unwanted interactions. In an ideal case with no spurious crosstalk, the decay rate measured in a simRB will match perfectly with that obtained in a single-qubit RB experiment. In Figs.~\ref{fig:SingleRB}(b,c) we see that the average gate error is slightly higher for simRB, which we ascribe to small classical correlations or  control parameter drifts~\citep{gambetta2012characterization}. Such correlations can arise from hardware crosstalk between the qubit control lines, specifically when the qubits are closely spaced in frequency as is the case in the measured device.
%
\section{\label{sec:level4}iSWAP gate Characterization}
%
%
The iSWAP gate is a two-qubit entangling operation. In superconducting qubits, it is typically implemented using tunable couplings to mediate excitation transfers between data qubits. To realize the two-qubit iSWAP operation, we employ parametric interactions between the data qubits, enabled by a fast flux modulation of the DTC. In this section, we first present the characterization of achievable parametric coupling magnitues used to resonantly activate the two-qubit exchange interaction. Then, we characterize the iSWAP gate using process tomography and two-qubit randomized benchmarking. 
%
\subsection{\label{sec:level4a}Parametric coupling}
%
\begin{figure}[b!]
     \centering
     \includegraphics[width=0.95\linewidth]{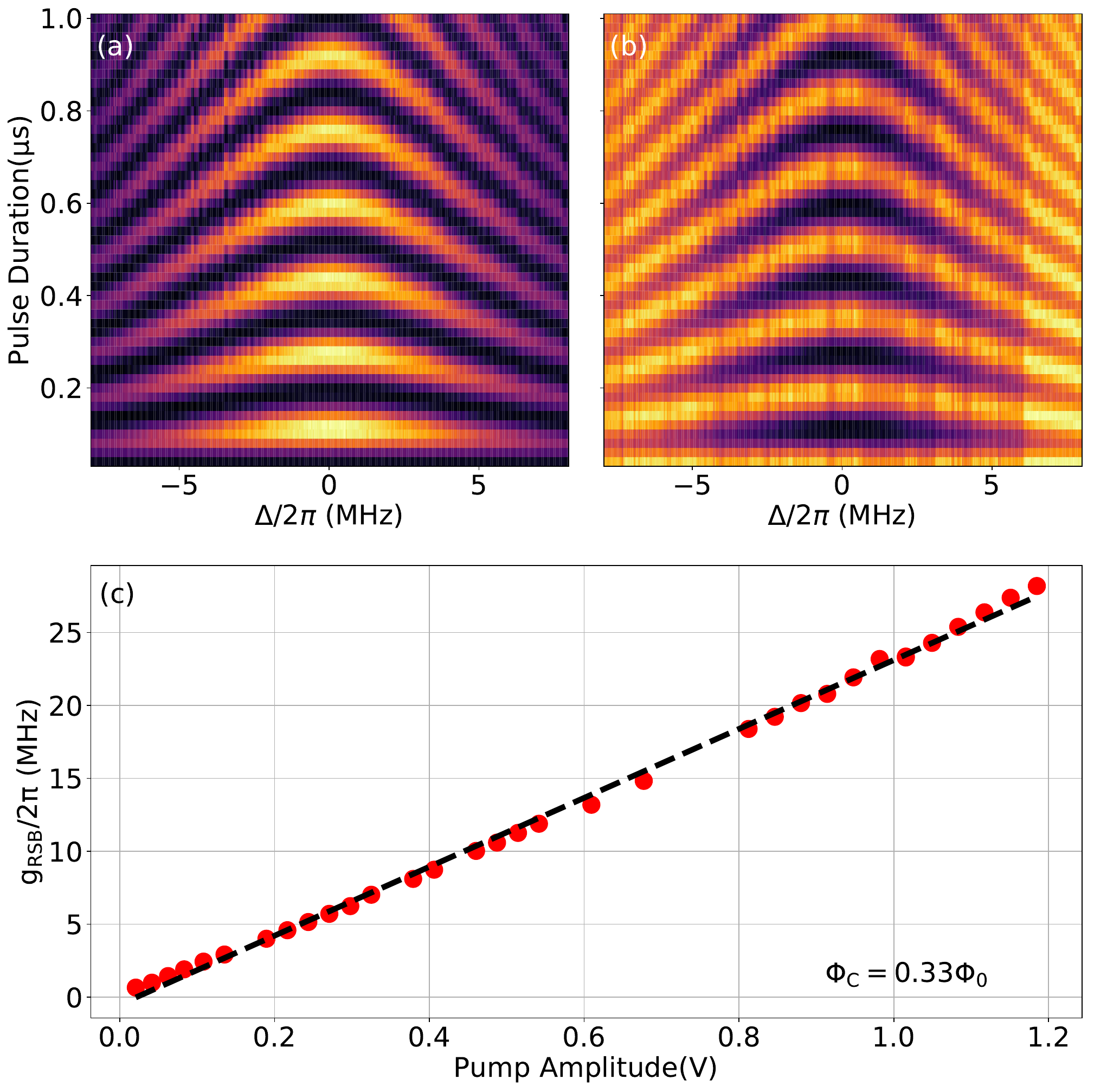}
     \caption{Parametric swaps measured as a function of pulse duration and detuning ($\Delta$) of the flux pump with respect to the difference frequency of data transmons ($\Delta_{\rm RSB}$). (a) Qubit $Q_1$ and (b) qubit $Q_2$. (c) Parametric coupling $g_{\rm RSB}$ inferred from the swap time, measured as a function of pump amplitude at room temperature.}
     \label{fig:RSB}
 \end{figure}
We perform a time-dependent modulation of the DTC flux using a high-frequency drive $\Phi(t)=\delta\Phi\cos(\omega_pt+\phi_p)$, which is fed through the RF port of the bias tee connected to the three-junction loop of the DTC [see Fig.~\ref{Fig:Fig1}(a)]. On tuning this drive in resonance with the difference (or `red sideband') of the two data qubit frequencies, $\omega_p=\omega_{2}-\omega_{1}$, it leads to a flux modulation of $g_{\rm eff}$ that mediates a parametric exchange interaction of magnitude \citep{zakka2011quantum},
\begin{equation}
g_{\rm RSB}=\frac{\partial g_{\rm eff}}{\partial \Phi}\delta\Phi\approx\frac{1}{2}\sqrt{\frac{\partial \omega_{1}}{\partial \Phi}\frac{\partial \omega_{2}}{\partial \Phi}}   \delta\Phi,
\end{equation}
where $\delta\Phi$ is the flux modulation amplitude at a given flux bias $\Phi_c$.  Since there are no static flux excursions required, the qubits are minimally detuned from the ZZ cancellation point during the gate, thus reaching high fidelity without any complicated pulse engineering. This is unlike resonant iSWAP gates, where the qubits are tuned into resonance with each other and away from their low $g_{\rm zz}$ point. In the latter case, special pulse envelopes are necessary to cancel any parasitic cross-Kerr interactions that may be active during the gate operation~\citep{sung2021realization}.

Detuning the parametric drive ($\Delta$) from `resonance' with the difference frequency of the data qubits ($\Delta_{RSB}$), reveals a chevron pattern showing the parametrically-induced excitation exchange [Fig.~\ref{fig:RSB}(a,b)]. The hopping rate $g_{\rm RSB}$ can be directly extracted from the frequency of these oscillations, which increases with the parametric drive amplitude. As shown in Fig.~\ref{fig:RSB}(c), the parametric coupling rate scales linearly with pump amplitude with a maximum achievable coupling corresponding to a 40\,ns swap rate, after which nonlinear mixing effects start to dominate.
%
\subsection{Robust phase estimation (RPE) for gate calibration}
\label{sec:level4b}
%
Traditional approaches for two-qubit gate calibration rely on numerical optimization of randomized benchmarking fidelity~\citep{jin2025superconducting,li2024realization} or tomography~\citep{sung2021realization}. However, these methods are only quadratically sensitive to calibration errors, which limits calibration accuracy. An alternative is robust phase estimation (RPE), which enables a high precision Heisenberg-limited measurement of gate errors~\citep{kimmel2015robust,rudinger2025heisenberg}. The general principle of RPE rests on amplifying the phase of the eigenvalues (a.k.a. eigenphases) of an unknown gate achieved by its repeated application: this typically involves preparing the qubits in a suitable superposition of two eigenvectors, applying the unknown gate repeatedly and then measuring in an appropriate basis to detect the sine and cosine of the relative eigenphase multiplied by the number of applied pulses. Recently, RPE was successfully used to calibrate a CZ gate on an 8-qubit transmon quantum processor~\citep{rudinger2025heisenberg}.  

Here we present a generalized RPE routine that can be used to estimate any unknown two-qubit gate parameter. We use this to devise a calibration procedure for the parametric iSWAP gate, that can deconvolve the errors caused by the phase and amplitude of the parametric pump and any Stark shifts on the individual qubits (see appendix~\ref{sec:appendix_iSWAPRPE}). It is worth noting that, unlike the case of CZ gates, the dominant error terms for the iSWAP operation do not commute, since $\sigma_{Zj}$ does not commute with $\sigma_{Xj}$, for $j=1,2$. As a result both eigenvectors and eigenvalues of the iSWAP matrix depend on the error terms, which renders standard routine for phase estimation insufficient for determining all the unknown gate errors. In order to calibrate the iSWAP we, therefore, build sequences of repeated compound gates that are designed to convert all unknown parameters into eigenphases that can then be efficiently amplified via RPE. 

To this end, we begin by parametrizing the iSWAP gate as~\citep{zhang2024tunable},  
\begin{align}
&{\rm iSWAP}(\theta_p,\phi_d,\theta_1,\theta_2,\phi_{\rm zz}) \nonumber\\
&= \begin{pmatrix}
    1 & 0 & 0 & 0\\
    0 & e^{i\theta_2}\cos(\theta_p) & ie^{i(\theta_2+\phi_p)}\sin(\theta_p) & 0\\
    0 & ie^{i(\theta_1-\phi_p)}\sin(\theta_p) & e^{i\theta_1}\cos(\theta_p) & 0\\
    0 & 0 & 0 & e^{i(\theta_1 + \theta_2+\phi_{zz})}\\
\end{pmatrix},
\label{eq:iswap}
\end{align}
where the angles $\theta_p$ and $\phi_p$ are set by the parametric pump amplitude and pump phase respectively, and $\theta_1$ and $\theta_2$ are induced by Stark shifts of the data qubits $Q_1$ and $Q_2$ respectively.
Various parameters in Eq.~(\ref{eq:iswap}) can be tuned experimentally, and $\theta_{1,2}$ can be corrected by virtual Z rotations. The error due to residual $g_{\rm zz}$ also appears as an additional rotation $\phi_{zz}$ on state $\ket{11}$. For the ideal iSWAP gate, $\theta_1=\theta_2=\phi_p=\phi_{zz}=0$ and $\theta_p=\pi/2$. Moreover, when $\theta_p\approx\pi/2$ and $\phi_{zz}\approx 0$, the number of remaining independent gate parameters in Eq.~(\ref{eq:iswap}) drops from three to two, since the sum of the phases of the off-diagonal terms $\theta_2+\phi_p$ and $\theta_1-\phi_p$ equals the phase of the last term. Therefore, without loss of generality, we set $\phi_p=0$ and calibrate the remaining parameters. Physically, this corresponds to the fact that applying equal and opposite Z rotations to each qubit frame is equivalent to shifting the parametric drive phase by twice the same amount. 

\begin{figure}[t!]
    \centering
    \includegraphics[width=\linewidth]{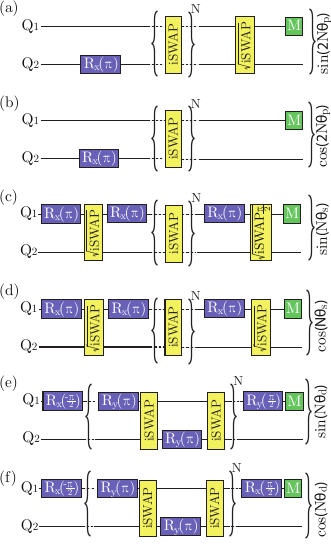}
    \caption{Pulse sequences used for iSWAP calibration. Panels (a)-(b) show the sequences used for amplification of rotation error due to pump amplitude ($\theta_p$), while (c)-(d) [(e)-(f)] show the sequences for amplification of the sum $\theta_s$ [difference $\theta_d$] of rotation errors induced by individual qubit Stark shifts. Here $\sqrt{\rm iSWAP_{\pi/2}}=\rm iSWAP(\pi/4,\pi/2,0,0,0)$, see Eq.~(\ref{eq:iswap}). The sequences amplify the unknown gate errors, but suppress state preparation and measurement (SPAM) errors at increasing $N$.}
    \label{fig:RPESeqs}
\end{figure}

The major difficulty in calibrating non-commuting error terms lies in the fact that the gate eigenvalues and eigenvectors are in general nonlinear functions of the unknown gate parameters, significantly complicating eigenphase estimation and necessitating a full process tomography or some other nonlinear optimization procedure. However, a closer look at eigenvalues of Eq.~(\ref{eq:iswap}) suggests that the problem can be greatly simplified since $\theta_p$ and the sum  $\theta_s=\theta_1+\theta_2$ are eigenphases of iSWAP, and thus can be amplified using standard phase estimation sequences consisting of repeated iSWAPs [Figs.~\ref{fig:RPESeqs}(a)-(d)]. The remaining task is to calibrate the difference $\theta_d=\theta_1-\theta_2$ which is accomplished by repeating the compound gate shown in Figs.~\ref{fig:RPESeqs}(e)-(f) with $\theta_d$ as an eigenphase, which consists of the repeated sequence ($Y_{Q_1},\rm{iSWAP},Y_{Q_2},\rm{iSWAP}$) with $Y_{Q_i}$ denoting a $\pi$ pulse on qubit $i$. Informed by these insights, we detail the full calibration procedure below:
\begin{itemize}
    \item Obtain a rough estimate of $\theta_p$ by fitting the center of the red sideband chevron measured in Fig.~\ref{fig:RSB}(a).
    \item Obtain a rough initial estimate of $\theta_{1,2}$ using the sequences in Fig.~\ref{fig:RPESeqs}(c,f) for measurement of sum $\theta_s=\theta_1+\theta_2$ and difference $\theta_d=\theta_1-\theta_2$ angles with $N=1$ repetitions.
    \item Use RPE to obtain an accurate value for $\theta_p$ by repetition of iSWAP and projection onto $|01\rangle$ and $|+\rangle=(|01\rangle+i|10\rangle)/\sqrt{2}$, see Figs.~\ref{fig:RPESeqs}(a,b).
    \item Measure $\theta_s$ by preparing $|\Psi\rangle=(|00\rangle+|11\rangle)\sqrt{2}$, repeating iSWAP and projecting onto $|\Psi\rangle$, see Figs.~\ref{fig:RPESeqs}(c,d).
    \item Use RPE on the compound gate described above to estimate the difference $\theta_d$, see Figs.~\ref{fig:RPESeqs}(e,f).
\end{itemize}

\begin{figure*}[t!]
    \centering
    \includegraphics[width=0.9\linewidth]{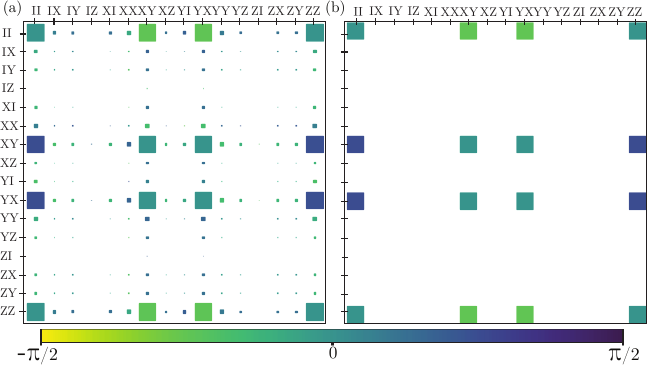}
    \caption{Comparison of process transfer matrices for the iSWAP gate, measured experimentally (a) and calculated theoretically (b). Size of the box represents the magnitude while color represents the phase of the respective matrix element. }
    \label{fig:ProcessTomo}
\end{figure*}

We observe that the sequences for amplification of $\theta_s$ and $\theta_p$ require a $\sqrt{\rm{iSWAP}}$ gate for state preparation and/or measurement. However, RPE is robust against SPAM errors, since the effect of SPAM errors is suppressed at increasing $N$. Therefore, we first approximately determine the parametric drive amplitude for iSWAP and then reduce the amplitude by half to implement $\sqrt{\rm{iSWAP}}$.

Interestingly, compared to the other gate parameters, the measurement of $\theta_d$ requires twice as many iSWAP pulses to produce the same amplification factor; therefore, the maximum number of repetitions needed for calibrating $\theta_d$ is  reduced by half. We typically use sequences of length $N=64$ to calibrate $\theta_s$ and $\theta_p$, with the maximum length of the calibration sequences being limited by qubit decoherence. Finally, in order to compensate for individual qubit Stark shifts, we correct the phase of the single-qubit drives by applying virtual Z rotations after iSWAP without performing any physical gate~\citep{mckay2017efficient}. 

Figure \ref{fig:ProcessTomo} shows the results of process tomography run at minimum $g_{\rm zz}$ operating flux with a 40\,ns iSWAP pulse. We estimate a gate fidelity of 99.827($\pm 0.004$)\%, with excellent agreement between the experimental data and theoretical prediction (see appendix~\ref{sec:appendix_processtomo}). From the measured iSWAP process matrix, we can extract an estimate of the nearest unitary gate by performing a Kraus decomposition,
\begin{equation}
\mathcal{E}(\rho) = \sum_i p_i\widehat{K}_i\rho\widehat{K}^\dagger_i,
\end{equation}
where the $\widehat{K}_j$ with the highest weight $p_j$  corresponds to the closest unitary representation of the quantum process. By comparing $\widehat{K}_j$ to the iSWAP unitary in Eq.~(\ref{eq:iswap}), we can extract $\phi_{zz}$ due to residual $g_{\rm zz}$ directly by looking at the last diagonal element of the matrix. We obtain $\phi_{zz}=-0.0119 \,(\pm0.0023)$~radians from the process tomography, which is fully consistent with the value of $g_{\rm zz}$ extracted from the JAZZ experiment (Sec.~\ref{sec:level3}) and a gate duration $\tau=$~40\,ns. Specifically, $H_{\rm zz}=g_{\rm zz}\sigma_{z1}\sigma_{z2}/4$, leading to $\phi_{zz}\approx (g_{\rm zz}/4)\tau \approx -0.013$ radians.

While process tomography is a standard benchmarking protocol, it is potentially limited by sensitivity to State Preparation and Measurement (SPAM) errors and exhibits poor scaling with system size. We therefore also performed two-qubit randomized benchmarking, which is insensitive to SPAM and gives a measure of the error accrued over a long sequence of gates, to further validate the measured gate fidelity and stability. 
%
\subsection{Randomized Benchmarking}
%
Two-qubit randomized benchmarking (RB) is a widely used protocol to measure average error rates for two-qubit gates with robustness against state preparation and measurement errors. In two-qubit RB, the qubits are prepared in their ground state and a sequence comprising random Clifford operations, selected from the two-qubit Clifford group, is applied~\citep{gaebler2012randomized}. This is followed by a recovery gate given by the inverse of the product of the preceding gates, after which the recovery probability (fidelity) is calculated by computing the ground state population as,
\begin{equation}
    P_{\rm 00} = \frac{1+ \langle \sigma_{Z1} \rangle + \langle \sigma_{Z2} \rangle +\langle \sigma_{Z1}\sigma_{Z2} \rangle}{4}.
\end{equation}
As before, the average error rate per Clifford ($r$) can be estimated by fitting the measured recovery probability as a function of sequence length (number of Cliffords) to Eq.~(\ref{RB_Fidelity}) with $d=4$. To implement all gates in the two-qubit Clifford group, we decomposed every Clifford operation as a product of single-qubit gates and a two-qubit operation \citep{corcoles2013process}. For this purpose, we use the iSWAP as our two-qubit gate primitive in conjunction with single-qubit rotations to generate both CNOT and SWAP gates (see appendix~\ref{sec:appendix_RBCliffords}). With the tensor product of single-qubit Cliffords, iSWAP, CNOT and SWAP, any element of the full two-qubit Clifford group can then be implemented.

\begin{figure}[t!]
    \centering
    \includegraphics[width=\linewidth]{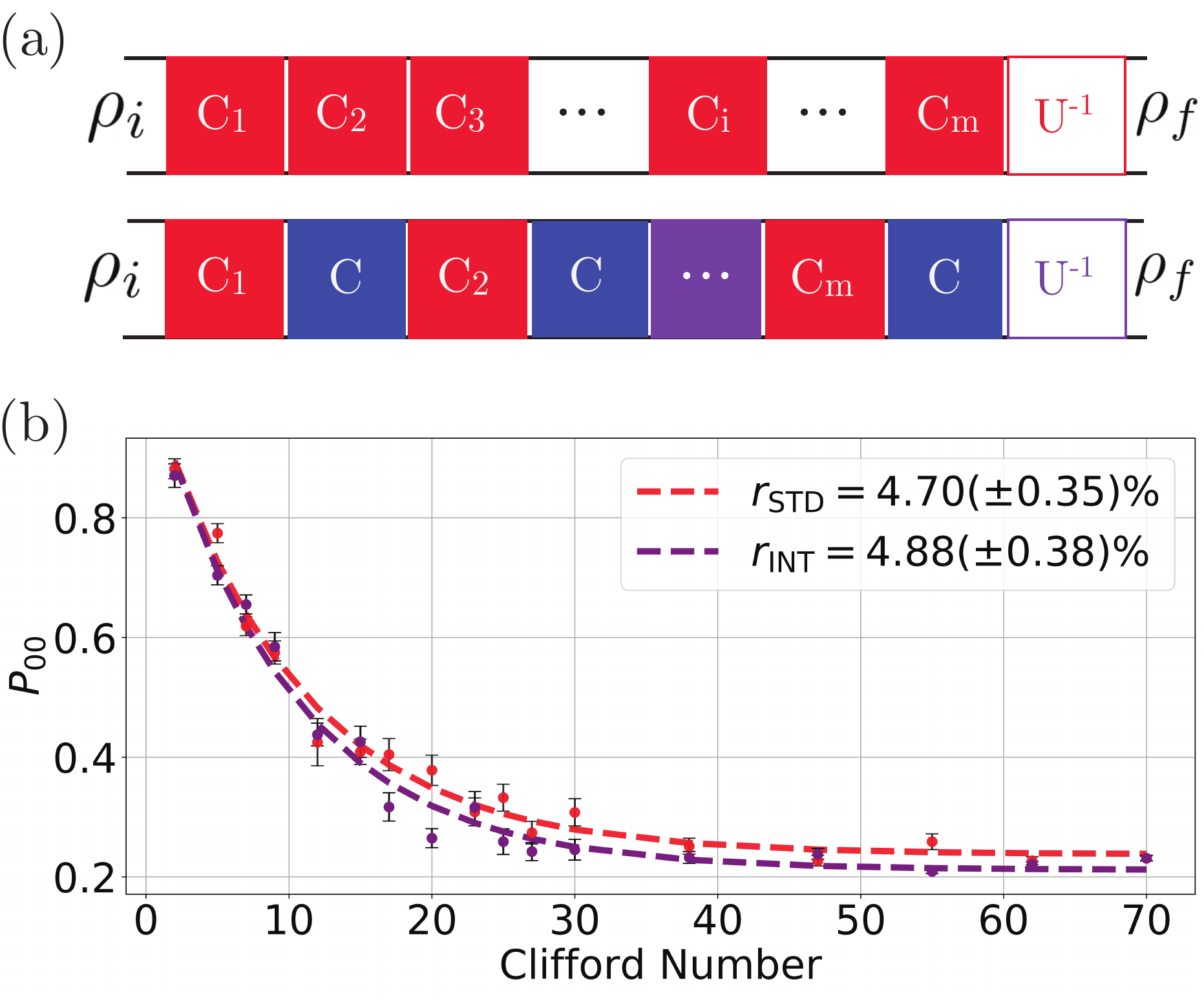}
    \caption{(a) Protocols for the standard and interleaved two-qubit randomized benchmarking (RB) for the iSWAP gate. The iSWAP gate is interleaved as a target (blue box). At the end, an inverse gate ($U^{-1}$) is added in order to render the overall operation an identity. (b) Experimental two-qubit RB with red (purple) circles showing the result of standard (interleaved) protocol. The respective error rates ($r_{\rm STD}$ and $r_{\rm INT}$) are extracted using an exponential fit to Eq.~(\ref{RB_Fidelity}).}
    \label{fig:TwoQubitRB}%
\end{figure}

The standard two-qubit RB provides the average gate fidelity of the full gate set. We estimate the fidelity of the iSWAP gate using interleaved two-qubit RB, where we insert an iSWAP before each random Clifford gate~\citep{magesan2012interleavedrb} (Fig. \ref{fig:TwoQubitRB}). Using the depolarizing parameters extracted from the standard and interleaved RB using Eq.~(\ref{RB_Fidelity}), we extract the fidelity of the iSWAP gate as,
\begin{equation} \label{RB_gate_fidelity}
    \mathcal{F}_{\rm iSWAP} = 1 - \frac{d-1}{d} \times \frac{p_{\rm STD}-p_{\rm INT}}{p_{\rm STD}},
\end{equation}
where $d=2^N$ and $N$ is the number of qubits in the system (here $N=2$), and $p_{\rm STD}$ and $p_{\rm INT}$ denote the depolarizing parameters obtained using standard and interleaved RB respectively. Using Eq.~(\ref{RB_gate_fidelity}), we extract an iSWAP gate fidelity of 99.70\%, which is consistent with the result obtained from process tomography. 

Notably, our measured gate fidelity is limited by qubit decoherence. We validate this conclusion by estimating the decoherence-limited fidelity based on a Markovian model of qubit decoherence during gate operation that gives, $F_{dec}=1-(2/5)\tau\sum_i\left(1/2T_1^i+1/T_2^i\right)$, where $T_1^i$, $T_2^i$ denote the relaxation and Hahn echo times of the two qubits respectively~\citep{abad2022universal}. Note that the use of Hahn echo times here is justified, since randomized benchmarking has an echoing effect that reduces low frequency noise, as also reported in other works~\citep{jin2025superconducting}. Using decoherence times of the data transmons, $T_1^1 =26.35 \,\mu s$, $T_1^2$ =17.0~$\mu s$, $T_2^1$=15.02~$\mu s$, $T_2^2$=17.11~$\mu s$, $\tau$=40\,ns, we estimate, ${F_{dec}\approx}$~99.72\%, while in the limit $T_2^i=2T_1^i$, the previous formula gives an upper decoherence limit ${F_{dec}\approx}$~99.84\%.

Evidently, the residual $g_{\rm zz}$ has limited impact on gate fidelity: using Eq.~(\ref{eq:iswap}), we estimate that the measured ${g_{\rm zz} = -2 \pi\times 220}$~kHz at the operating flux bias results in a gate error of $0.002\%$, well below the decoherence limit in our device. Previous works using DTC have demonstrated $T_2^{\rm echo}>$ 100 $\mu s$~\citep{li2024realization}, which can further serve to increase the achievable fidelity to above 99.9\% in such coupling architectures. 

\section{Conclusions}
\label{sec:level5}
%
In this work, we have demonstrated a high-fidelity parametric iSWAP gate between transmon qubits with 99.7\% fidelity in 40\,ns. We used a double transmon coupler architecture to statically decouple the qubits and suppress cross-Kerr interactions, as confirmed by the gate fidelity measurements which are shown to be entirely limited by the decoherence of data-qubits. Crucially, the bias flux for static coupling cancellation is internally defined by the DTC parameters making it robust against fabrication variations; further, this allows modular circuit layouts without the need for any coupler redesign when deployed to couple modes with a wide spread in frequencies. 

Since iSWAP operations do not commute with experimental calibration errors due to a.c.~Stark shifts and spurious cross-Kerr interactions, we developed a phase estimation technique which enables an efficient and high-precision measurement of all unknown gate parameters without relying on complicated pulse optimization, randomized benchmarking or full tomographic reconstruction. Our approach is directly extensible to other qubit platforms and two-qubit gates. 
%
\section{Acknowledgments}
%
We thank Kenneth Rudinger, Kevin Young and Xiaoyue Y. Jin for useful discussions. This material is based upon work supported by the U.S. Department of Energy, Office of Science under award number DE-SC0019461. The TWPA used in this experiment is provided by IARPA and MIT Lincoln Labs. Devices for this work were fabricated at both BBN and Rigetti quantum Foundry Services. We would like to thank Yuvraj Mohan and the Rigetti QFS team for design support and device fabrication. Any opinions, findings, and conclusions or recommendations expressed in this article are those of the authors and do not necessarily reflect the views of the Air Force Research Laboratory (AFRL). Approved for Public Release; Distribution Unlimited: PA\# AFRL-2026-2004.
%
\appendix
%
\section{Experimental Setup}
\label{sec:appendix_setup}
%
\begin{figure}[t!]
    \centering
    \includegraphics[width=\linewidth]{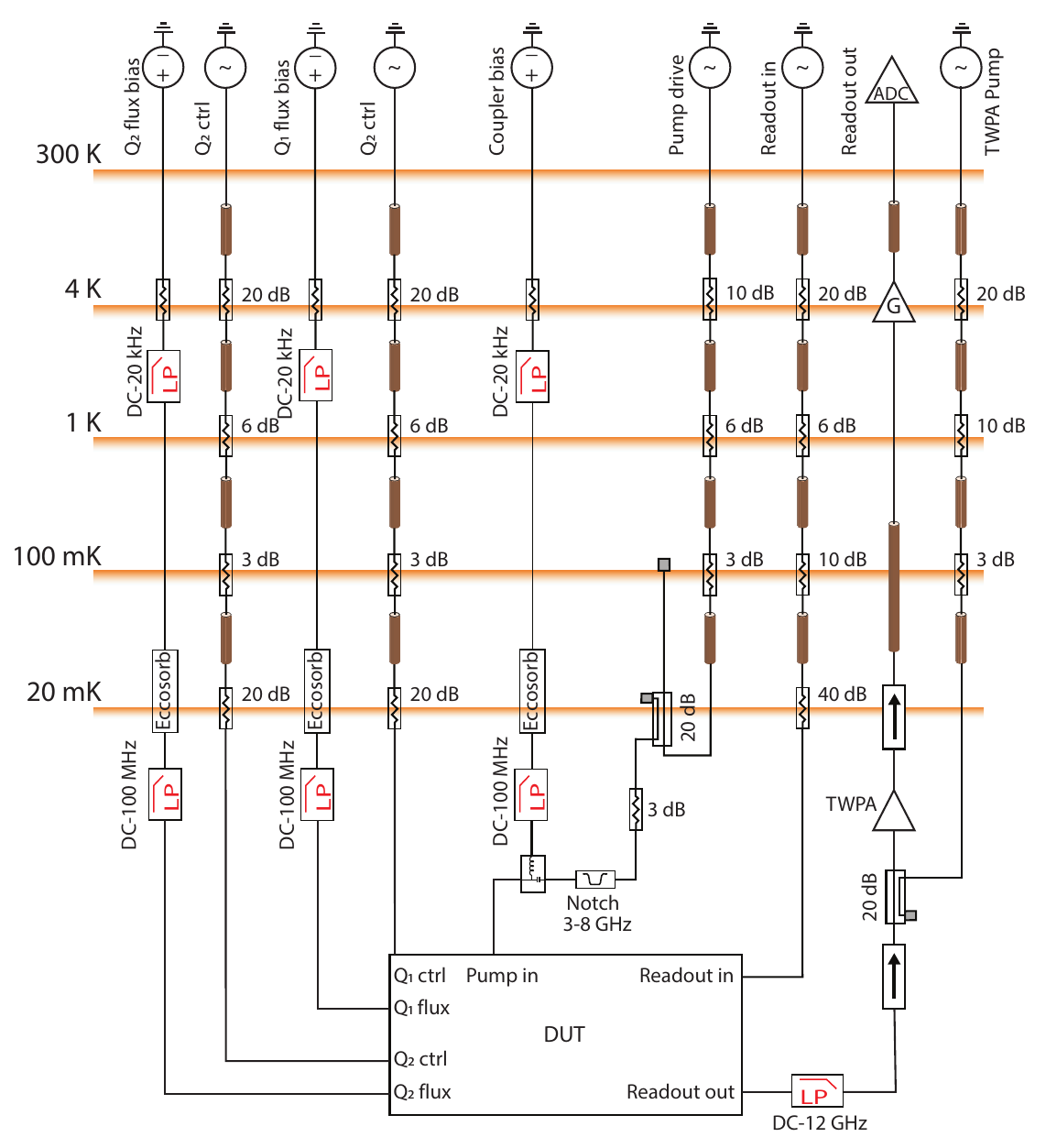}
    \caption{Experimental setup.}
    \label{fig:ExpSetup}
\end{figure}
The experimental setup is shown in Fig.~\ref{fig:ExpSetup}. Standard cryogenic attenuators are placed along the input RF lines, with a total of 76~dB attenuation along the readout line and 49~dB attenuation along the qubit control lines. The attenuation along the parametric pump lines is reduced to meet the higher power requirement for fast modulation. Additionally, we terminate the qubit parametric drive line into a 20~dB directional coupler to avoid excessive power dissipation at the mixing chamber. We placed a 3-8~GHz notch filter before the coupler flux port to protect the qubits from decay and an external bias tee to combine DC and RF flux drives. The output readout signal is amplified using a traveling wave parametric amplifier (TWPA), followed by a HEMT amplifier at 4K and room temperature amplification. We use DC-20~kHz resistive low-pass filters at the 4~K stage to filter noise in the flux bias lines followed by further low-pass filtering at the mixing chamber (Eccosorb and LC 100~MHz lowpass filtering).  
We generate the pump and qubit control signals up to 2~GHz using a Xilinx XCKU115 UltraScale FPGA and AD9164 DACs from Analog Devices~\citep{kalfus2020high}. The signals are converted to the qubit frequencies using an external mixer. The readout tones are modulated using fast waveform generators (Advanced Pulse Sequencer)~\citep{ryan2017hardware}. We apply two different modulation frequencies and combine the resulting waveforms to perform joint qubit readout. 
%
\section{Circuit simulation and static cross-Kerr cancellation}
\label{sec:appendix_zz}
%
According to the analytical description of the double transmon coupler~\citep{campbell2023modular,goto2022double}, we expect to achieve a perfect cancellation of $g_{\rm zz}$. However, in our experimental device we observed a residual interaction of 220~kHz, as detailed in the main text. We attribute the discrepancy to the presence of a second-order, $\mathcal{O}(g_{\rm eff}^{2})$, cross-Kerr contributions by the higher energy levels of the coupler. To confirm this, we simulated the circuit shown in Fig.~\ref{Fig:Fig1} using the following Hamiltonian in scQubits~\citep{groszkowski2021scqubits},
\begin{widetext}
\begin{align}
H &= \sum_{i=1...4}4E_{Ci}(\hat{n}_i-n_{gi})^2-E_{J1}\cos(\hat{\phi_1})-E_{J2}\cos(\hat{\phi}_2)-E_{J3}\cos(\hat{\phi}_3+\phi_e)-E_{J4}\cos(\hat{\phi}_4)-E_{J5}\cos(\hat{\phi}_3-\hat{\phi}_4) \nonumber\\
& \qquad +4E_{C13}(\hat{n}_1-n_{g1}-\hat{n}_3+n_{g3})^2+4E_{C24}(\hat{n}_2-n_{g2}-\hat{n}_4+n_{g4})^2+4E_{C34}(\hat{n}_3-n_{g3}-\hat{n}_4+n_{g4})^2,
\end{align}
\end{widetext}
where $E_{Cjk}=e^2/2C_{jk}$ denote the charging energies associated with the capacitors $C_{jk}$. $E_{Ji}$ denote the ith-junction Josephson energy. Note that for static simulations the external flux $\phi_e$ can be assigned to any of the coupler junctions (we chose $\phi_3$), unlike time-dependent simulations which may require an appropriate gauge choice~\citep{you2019circuit,bryon2023time,riwar2022circuit,campbell2023modular}.  

\begin{figure}[b!]
    \includegraphics[width=\linewidth]{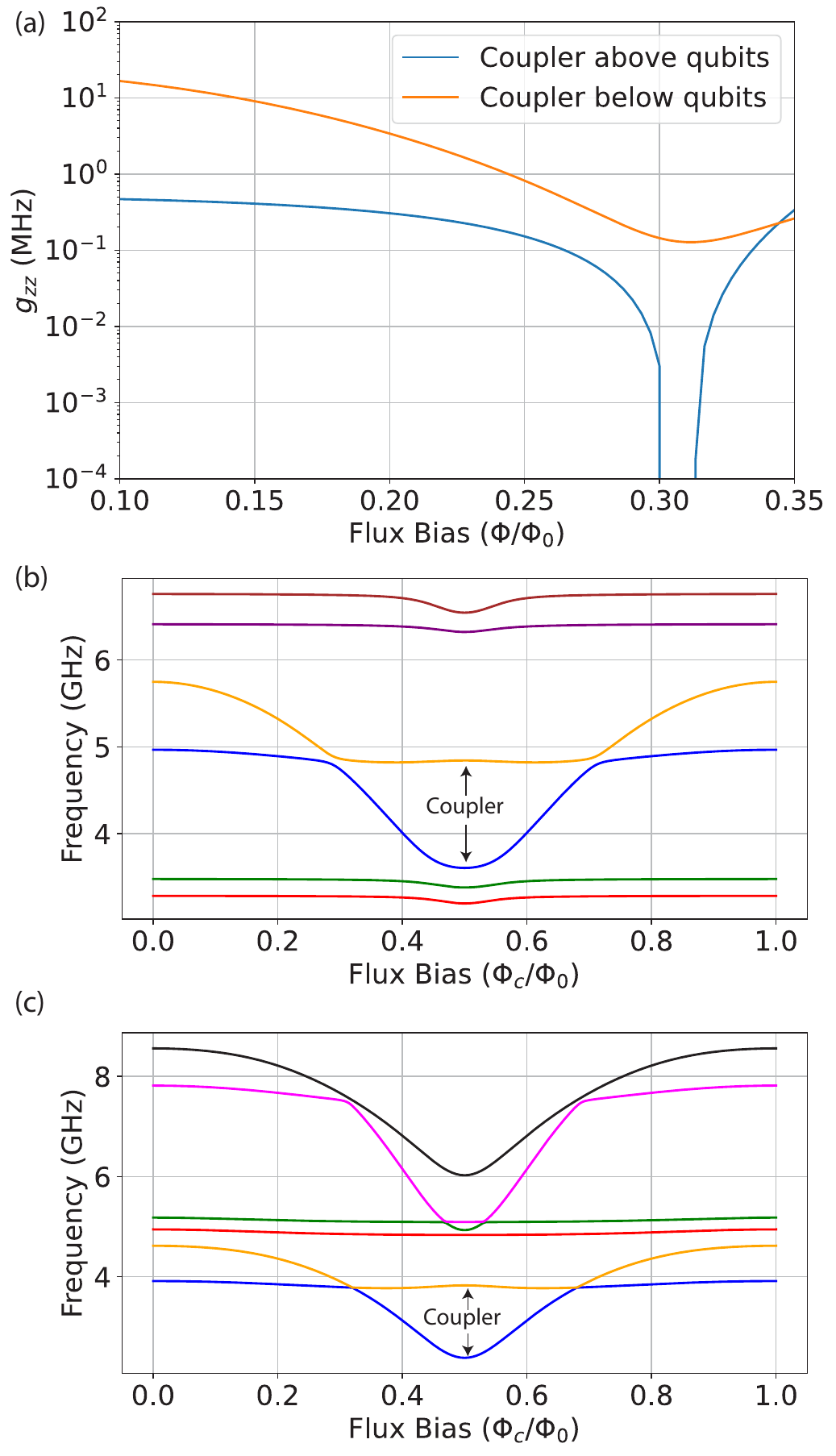}
    \caption{(a) Comparison of simulated cross-Kerr interaction between two transmons coupled via the DTC with the coupler placed above (blue) or below (orange) the data qubits. We expect perfect cancellation when the coupler frequency is higher than the data-qubit frequencies, because of reduced participation of the higher energy levels of the coupler. Panel (b) [(c)] shows the lowest six energy levels when the coupler is parked below (above) the data qubits.}
    \label{fig:ZZSim}
\end{figure}

We performed the simulation for two cases, namely, when the coupler frequency is placed above or below the data qubit frequencies. The results of the simulations are summarized in Fig.~\ref{fig:ZZSim}. In Fig.~\ref{fig:ZZSim}(a), the simulated cross-Kerr interaction $g_{\rm zz}$ shows a perfect cancellation  near $\Phi=0.3\Phi_0$ when the coupler is above the qubits (blue line). As seen in Fig.~\ref{fig:ZZSim}(b), in this situation the coupler transmons can be accurately treated as two-level systems and the analytical theory fully applies. On the contrary, when the coupler frequency is lower than that of the data qubits, we observe a minimum $g_{\rm zz}$ greater than 100~kHz (Fig.~\ref{fig:ZZSim}(a), orange line). This corresponds to the case where the second-excited states of the coupler qubits interact strongly with the data qubits near the cancellation flux [see Fig.~\ref{fig:ZZSim}(c)] leading to an additional cross-Kerr correction. We also report the circuit parameters for the device measured and discussed in the main text, which correspond to the simulation in Fig.~\ref{fig:ZZ_cancellation}(c), in Table~\ref{tab:device_params}.

\begin{table}
    \centering
    \begin{tabular}{l|c}
        {\bf Circuit Parameter} & {\bf Value} \\
        \hline
        $Q_1$ junction energy $E_{J1}/h$ & 19.5~GHz   \\
        \hline
        $Q_2$ junction energy $E_{J2}/h$&  20.0~GHz \\
        \hline
        $Q_{3,4}$ junction energy $E_{J3,4}/h$ & 14.6~GHz\\
        \hline
        $Q_1$ anharmonicity $E_{C1}/h$ & $-164$~MHz   \\
        \hline
        $Q_2$ anharmonicity $E_{C2}/h$&  $-172$~MHz \\
        \hline
        $Q_{3,4}$ anharmonicity $E_{C3,4}/h$ & $-150$~MHz\\
        \hline
        $J_5$ junction energy $E_{J5}/h$ & 5.8~GHz \\
        \hline
         $Q_1$-$Q_3$ coupling capacitance $C_{13}$ & 13.5~fF \\
        \hline
         $Q_2$-$Q_4$ coupling capacitance $C_{24}$ & 13.9~fF \\
        \hline
        $Q_3$-$Q_4$ coupling capacitance $C_{34}$ & 7.91~fF \\
        \hline
        $Q_1$ readout resonator coupling capacitance $C_{1R}$ & 4.0~fF \\
        \hline
         $Q_2$ readout resonator coupling capacitance $C_{2R}$ & 3.5~fF \\
        \hline
         $Q_3$ readout resonator coupling capacitance $C_{3R}$ & 7.5~fF \\
        \hline
         $Q_4$ readout resonator coupling capacitance $C_{4R}$ & 7.5~fF \\
        \hline
         $Q_{1,2}$ control line coupling capacitance $C_{1Q,2Q}$ & 0.06~fF \\
        \hline
    \end{tabular}
    \caption{Nominal device parameters for the circuit shown in Fig.~\ref{Fig:Fig1}(a).}
    \label{tab:device_params}
\end{table}
%
\section{iSWAP Phase estimation} 
\label{sec:appendix_iSWAPRPE}
%
As discussed in the main text, calibration of the iSWAP gate is non-trivial because of the non-commuting nature of the error terms. We use Robust Phase Estimation (RPE) to amplify the error terms by repeating the unknown gate. Error amplification requires three main steps: (i) Preparation of the qubits in a suitable initial state, (ii) cascade of either the iSWAP or a suitably constructed compound gate, and (iii) final measurement in a suitable basis. 

Let us assume we want to estimate the eigenvalues of an unknown unitary $\hat{U}$, represented using its spectral decomposition, 
\begin{equation}
    \hat{U} = \sum_i e^{i\theta_i}|\Psi_i\rangle \langle \Psi_i|,
\end{equation}
where, $\theta_i$, $|\Psi_i\rangle$ index the eigenphases and eigenvectors of $\hat{U}$. We note that RPE can only capture phase difference of eigenphases, since any arbitrary phase prefactor is not observable. In order to estimate the relative phase $\theta_{ab} = \theta_a - \theta_b$, corresponding to the eigenstates $|\Psi_a \rangle$ and $|\Psi_b \rangle$, we initialize the system into the following state,
\begin{align}
    |\Psi_{in} \rangle &= \frac{|\Psi_a \rangle + |\Psi_b \rangle}{\sqrt{2}}.
\end{align}
We then amplify $\theta_{ab}$ by applying the desired unitary $N$ times,
\begin{equation} \label{state_amp}
    |\Psi_S \rangle = \hat{U}^N |\Psi_{in} \rangle = \frac{e^{iN\theta_a}|\Psi_a \rangle + e^{iN\theta_b}|\Psi_b \rangle}{\sqrt{2}}.
\end{equation}
Finally, we need to project the state onto $\ket{\Psi_{a,b}}$ in order to measure the sine and cosine of $N(\theta_b-\theta_a)$. This is achieved by applying a suitable unitary ($\hat{U}^{c,s}_M$) before the final measurement in the Z basis. In general, we will need to measure only one qubit, which we assume to be $Q_1$. The measurement operator ($\hat{O}_M$) is then equal to,
\begin{equation}
    \hat{O}_M = \left (\hat{U}^{c,s}_M \right)^{\dagger} \left (\sigma_z \otimes I \right ) \hat{U}^{c,s}_M.
\label{eq:OM}
\end{equation}
Here the superscripts $\{c,s\}$ identify the measurement operators applied for measuring the cosine and sine of the estimated angle respectively. We estimate $n\theta_{ab}$ for $n=1 \dots N$ from the sine and cosine measurements from the sequence of length $n$. For each value of $n$ there are $n+1$ possible values of $\theta_{ab}$ consistent with the measurements. According to the standard RPE algorithm~\citep{kimmel2015robust,rudinger2025heisenberg}, for each value of $n$, we select the angle closest to the estimate at $n-1$ resulting in an increasingly accurate estimate with increasing $n$.

As an example, we show the sequences to estimate $\theta_p$ in Fig.~\ref{fig:RPESeqs}(a,b). We verified that the calibration converges faster if we estimate $\theta_p$ before $\theta_{1,2}$. Before proceeding with phase estimation, we perform an initial rough fit of the gate parameters via the sequences in Fig.~\ref{fig:RPESeqs} for $N=1$. For small errors, we can analyze the calibration of $\theta_p$ with $\theta_1=\theta_2=\phi_p\approx0$, with the following form of the iSWAP matrix,
\begin{align} \label{casc_iswap_p}
\rm iSWAP(\theta_p) = \begin{pmatrix}
    1 & 0 & 0 & 0\\
    0 & \cos(\theta_p) & i\sin(\theta_p) & 0\\
    0 & i\sin(\theta_p) & \cos(\theta_p) & 0\\
    0 & 0 & 0 & 1\\
\end{pmatrix},
\end{align}
corresponding to the following eigenvalues and eigenvectors, 
\begin{center}
    \begin{tabularx}{0.92\columnwidth}{l|l}
    {\bf Eigenvalues} & {\bf Eigenvectors} \\
    \hline
    1 & $|\psi_1\rangle=(1,0,0,0)^{T}$  \\
    \hline
    $\exp(-i\theta_p)$ &$|\psi_2\rangle=2^{-1/2}(0,-1,1,0)^{T}$  \\
    \hline
    $\exp(i\theta_p)$ & $|\psi_3\rangle=2^{-1/2}(0,1,1,0)^{T}$  \\
    \hline
    1 & $|\psi_4\rangle=(0,0,0,1)^{T}$  \\
    \hline
    \end{tabularx}
\end{center}
Following the recipe introduced above, we can measure $\theta_p$ by initializing the two-qubit system in the state,
\begin{equation}
    |\Psi (0) \rangle = \frac{|\psi_2\rangle + |\psi_3\rangle}{\sqrt{2}}=|01\rangle,
\end{equation}
by applying a $\pi$ pulse on $Q_b$ [Fig.~\ref{fig:RPESeqs}(a,b)]. Then, the iSWAP gate is applied $N$ times ($\hat{U}^N$) leading to the state,
\begin{equation}
    |\Psi_S \rangle = \hat{U}^N|\Psi(0) \rangle = \frac{e^{-iN\theta_p}|\psi_2\rangle + e^{iN\theta_p}|\psi_3\rangle}{\sqrt{2}}.
\end{equation}
For the final measurement, we apply $\hat{U}^s_M = \sqrt{\rm iSWAP}$ and measure $Q_a$, as in Fig.~\ref{fig:RPESeqs}(a), resulting in the measurement operator from Eq.~(\ref{eq:OM}),
\begin{align} 
\hat{O}_M =  \begin{pmatrix}
    1 & 0 & 0 & 0\\
    0 & 0 & i & 0\\
    0 & -i & 0 & 0\\
    0 & 0 & 0 & -1\\
\end{pmatrix}
\end{align}
and the final measurement 
\begin{equation}
    \langle \Psi_S |\hat{O}_{M} | \Psi_S \rangle = \sin(2N\theta_p).
\end{equation}
Similarly, we choose $\hat{U}^c_M = I$ to measure $\cos(2N\theta_p)$ as in Fig.~\ref{fig:RPESeqs}(b). After calibrating $\theta_p$, we can measure the remaining gate parameters in the following model,
\begin{align} \label{casc_iswap_st}
\rm iSWAP (\theta_1,\theta_2) = \begin{pmatrix}
    1 & 0 & 0 & 0\\
    0 & 0 & e^{\theta_2} & 0\\
    0 & e^{\theta_1} & 0 & 0\\
    0 & 0 & 0 & e^{\theta_1+\theta_2}\\
\end{pmatrix}.
\end{align}
In Eq.~(\ref{casc_iswap_st}) we assumed $\phi_p=0$, since $\phi_p$ can be equivalently included in $\theta_2-\theta_1$ by a simple reparameterization of the gate matrix. This is equivalent to a frame rotation of the qubits and pump, which has no physically measurable effect.

The eigensystem of the matrix in Eq.~(\ref{casc_iswap_st}) is
\begin{center}
    \begin{tabularx}{0.92\columnwidth}{l|l}
    {\bf Eigenvalues} & {\bf Eigenvectors} \\
    \hline
    1 & $|\psi_1\rangle=(1,0,0,0)^{T}$  \\
    \hline
    $-i\exp(i\theta_s/2)$ &$|\psi_2\rangle=2^{-1/2}(0,-\exp(-i\theta_d/2),1,0)^{T}$  \\
    \hline
    $i\exp(i\theta_s/2)$ & $|\psi_3\rangle=2^{-1/2}(0,\exp(i\theta_d/2),1,0)^{T}$  \\
    \hline
    $\exp(i\theta_s/2)$ & $|\psi_4\rangle=(0,0,0,1)^{T}$  \\
    \hline
    \end{tabularx}
\end{center}
where $\theta_{s} = \theta_{1} + \theta_{2}$ denotes the sum of single-qubit rotation errors. Clearly, to calibrate $\theta_s$, the initial set of pulses must combine the $|00\rangle$ and $|11\rangle$ states. Therefore, the initial pulses in Fig. \ref{fig:RPESeqs}(c,d) prepare the state 
\begin{equation}
    |\Psi (0) \rangle = \frac{|\psi_1\rangle + i|\psi_4\rangle}{\sqrt{2}} =\frac{|00\rangle + i|11\rangle}{\sqrt{2}}.
\end{equation}
After $N$ repetitions of $\hat{U}=\textrm{iSWAP}(\theta_1,\theta_2)$, the resultant system state is,
\begin{equation}
    |\Psi_S \rangle = \hat{U}^N|\Psi(0) \rangle = \frac{|00\rangle + ie^{iN\theta_s}|11\rangle}{\sqrt{2}}.
\end{equation}
For Fig.~\ref{fig:RPESeqs}(c), ${\hat{U}^{s}_M=\sqrt{\textrm{iSWAP}_{\pi/2}}=\textrm{iSWAP}(\pi/4,\pi/2,0,0)}$ [Eq.~(\ref{eq:iswap})]. Hence the parametric drive phase of the final gate before measurements is set to $\phi_p=\pi/2$ so that $\langle \Psi_S |\hat{O}_M | \Psi_S \rangle = \sin(N\theta_s)$. 
\par
To understand why compounding is necessary, it is important to notice that the information of $\theta_d  = \theta_1-\theta_2$ is encoded in the eigenvectors of Eq.~(\ref{casc_iswap_st}). From the matrix representation of the compound gate,
\begin{align} \label{compound_iswap}
\hat{U}_{cmp} &= Y_{Q1}({\rm iSWAP})Y_{Q2}({\rm iSWAP}) \nonumber\\
&=\begin{pmatrix}
    e^{i\theta_2} & 0 & 0 & 0\\
    0 & e^{i(\theta_2+2\theta_2)} & 0& 0\\
    0 & 0 & e^{i\theta_1} & 0\\
    0 & 0 & 0 & e^{2\theta_1+\theta_2}\\
\end{pmatrix},
\end{align}
we see that $\theta_d$ can be extracted as a relative phase by initializing the system in a superposition of $|00\rangle$ and $|10\rangle$. The initial set of pulses in Figs.~\ref{fig:RPESeqs}(e,f) prepares the state
\begin{equation}
    |\Psi (0) \rangle = \frac{|00\rangle + i|10\rangle}{\sqrt{2}}.
\end{equation}
The compound gate is applied $N$ times to this initial state which results in the following two-qubit state,
\begin{equation}
    |\Psi_S \rangle = \hat{U}_{cmp}^N|\Psi (0) \rangle = \frac{e^{iN\theta_2}|00\rangle + ie^{iN\theta_1}|10\rangle}{\sqrt{2}}.
\end{equation}
The measurement operators can be defined as per Figs.~\ref{fig:RPESeqs}(e,f) to measure the required trigonometric function.
\par
\begin{figure}[t!]
    \centering
    \includegraphics[width=\linewidth]{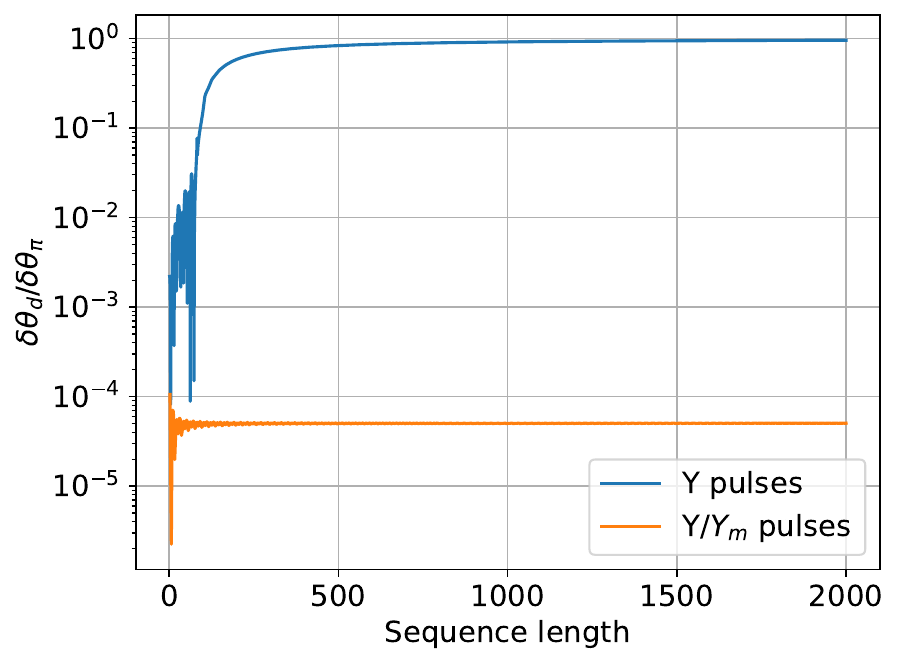}
    \caption{Propagation of single-qubit gate errors into iSWAP gate calibration errors for the sequences shown in Figs.~\ref{fig:RPESeqs}(e,f). The y-axis shows the ratio between the iSWAP angle calibration error $\delta\theta_d$ and the single-qubit $\pi$ pulse calibration error $\delta\theta_\pi$. For odd $N$ we insert $Y$ gates, while for even $N$ we insert $Y_m=ZYZ$ gates. By alternating between $Y$ and $Ym$ qubit rotations, we can cancel the effect of single-qubit gate errors on the iSWAP gate calibration.}
    \label{fig:Y_Ym_alt}
\end{figure}

One potential problem with using compound gates is that single-qubit rotation errors can accumulate and reduce the calibration accuracy of the iSWAP gate. Here we make the calibration of $\theta_d$ robust against single-qubit rotation errors by alternating between the gates $Y$ and $Y_m=ZYZ$ as the calibration sequence is built. This reduces any single-qubit gate errors which can accumulate as the compound iSWAP gate is applied. Figure~\ref{fig:Y_Ym_alt} simulates the effect of alternating between $Y$ and $Y_m$ pulses. In the simulation we assumed no decoherence and that all other gate parameters were calibrated.

The RPE routine can then be used to estimate the unknown relative eigenphase ($\theta$) with $N=2^k$ as the number of pulse repetitions by successive approximations. The precise value of $\theta$ is calculated using
\begin{equation} \label{Eq:phase_calc}
    \theta^{(k)} = \frac{1}{2^k} \Big( \textrm{arctan} \big(\sin(2^k \theta)/\cos(2^k \theta) \big) + 2\pi n_k\Big).
\end{equation}
The $n_k$ is chosen to ensure that $\theta^{(k)}$ and its predecessor $\theta^{(k-1)}$ are in the same quadrant. This restricts the eigenphase in the range $[\bar{\theta}-\pi/2^k,\bar{\theta}+\pi/2^k]$. The accuracy of phase estimation, therefore, depends on the coherence of two qubits. The RPE formulation also ensures that the eigenphase estimation error is Heisenberg-limited, i.e., $\sigma(\theta) \sim \mathcal{O}(1/2^k)$. 
%
\section{Process Tomography} 
\label{sec:appendix_processtomo} 
%
Quantum process tomography (QPT) allows to reconstruct a completely-positive and trace-preserving (CPTP) map that best describes an unknown quantum operation. A CPTP map of an unknown process can be represented as a linear map on the density matrix ($\rho$), using the process matrix $\chi_{i,j}$ matrix,
\begin{equation}
    \rho \rightarrow \sum_{i,j} \chi_{i,j} P_i\rho P_j,
\end{equation}
where each element of $\{P_i \}$ is an element of the $N$-qubit Pauli group $\{I,X,Y,Z\}^{\otimes N}$. The goal of QPT is to construct the $4^N\times4^N$ $\chi_{i,j}$ matrix, where $N$ is the number of qubits.

When performing QPT of a two-qubit gate, we need to reconstruct a $16 \times 16$ unknown process matrix. We prepare each qubit in 6 orthogonal initial states: $|0 \rangle,\frac{1}{\sqrt{2}}(|0\rangle \pm |1\rangle),\frac{1}{\sqrt{2}}(|0\rangle \pm i|1\rangle)$ and $|1\rangle$ by applying the following set of pulses:
 \begin{equation} \label{input_st}
     \{ P_{i} \} = \{I,R_x(\pm\pi/2),R_y(\pm\pi/2),R_x(\pi)\}^{\otimes 2},
 \end{equation} 
where,
\begin{equation}
     R_{\hat{n}}(\theta) = \exp[-i \frac{\theta}{2} (\hat{n}.\overrightarrow{\sigma}_n)].
\end{equation}
Here, the vector $\hat{n}$ is the axis of rotation on the Bloch sphere and $\overrightarrow{\sigma}_n=\{I,\sigma_x,\sigma_y,\sigma_z\}$ is the Pauli matrix vector. Thus, for a total of 36 orthogonal initial states, this results in an overdetermined set of equations. 

The unknown gate is then applied on each of the input states, followed by measurement of operators defined by the set in Eq.~(\ref{input_st}). This step is equivalent to performing a quantum state tomography for each of the gate output state for each of the 16 input states, leading to a total of 256 measurement configurations which is enough to reconstruct the unknown process matrix. For each of the configurations, both single-qubit ($\langle \sigma_{Z1} \rangle,\langle \sigma_{Z2} \rangle$) and two-qubit ($\langle \sigma_{Z1}\sigma_{Z2} \rangle$) operators are measured. The measured data are then compared to the expected result for the unknown process matrix and we use a least-squares fit to construct the process matrix $\chi_{i,j}$. We further add constraints to the least-squares fit algorithm to ensure that the resulting matrix is completely positive and trace preserving. Finally, we use bootstrap resampling to compute the 95\% confidence intervals for the reconstructed process matrix and extracted gate fidelity~\citep{home2009complete}. In bootstrap resampling, we use the mean and variance of the tomographic measurement records to generate 100 sample datasets. We used each dataset to perform a new tomographic inversion and obtain a statistical distribution of the process matrix entries. From this data we extracted confidence intervals of the relevant quantities. We visualize the $\chi_{i,j}$ as a Hinton plot as shown in Fig.~\ref{fig:ProcessTomo}. 
\par
After performing process tomography, we compare the measured process matrix to the theoretical iSWAP process matrix and calculate process fidelity:
\begin{equation}
    \mathcal{F}_{\textrm{process}} = \rm Tr (\chi_{\rm th} \chi_{\rm exp}),
\end{equation}
where, $\chi_{\rm th}$ and $\chi_{\rm exp}$ are the theoretically and experimentally derived process matrices respectively. The process fidelity can be used to calculate the gate fidelity $\mathcal{F}_{\textrm{gate}}$ as
\begin{equation}
    \mathcal{F}_{\textrm{gate}} = \frac{1+4\mathcal{F}_{\textrm{process}}}{5},
\end{equation}
which is valid for two-qubit operations with ${\rm dim}(\mathcal{H})=4$.
%
\section{Two-qubit Randomized Benchmarking} 
\label{sec:appendix_RBCliffords}
%
\begin{figure}[b!]
    \centering
    \includegraphics[width=\linewidth]{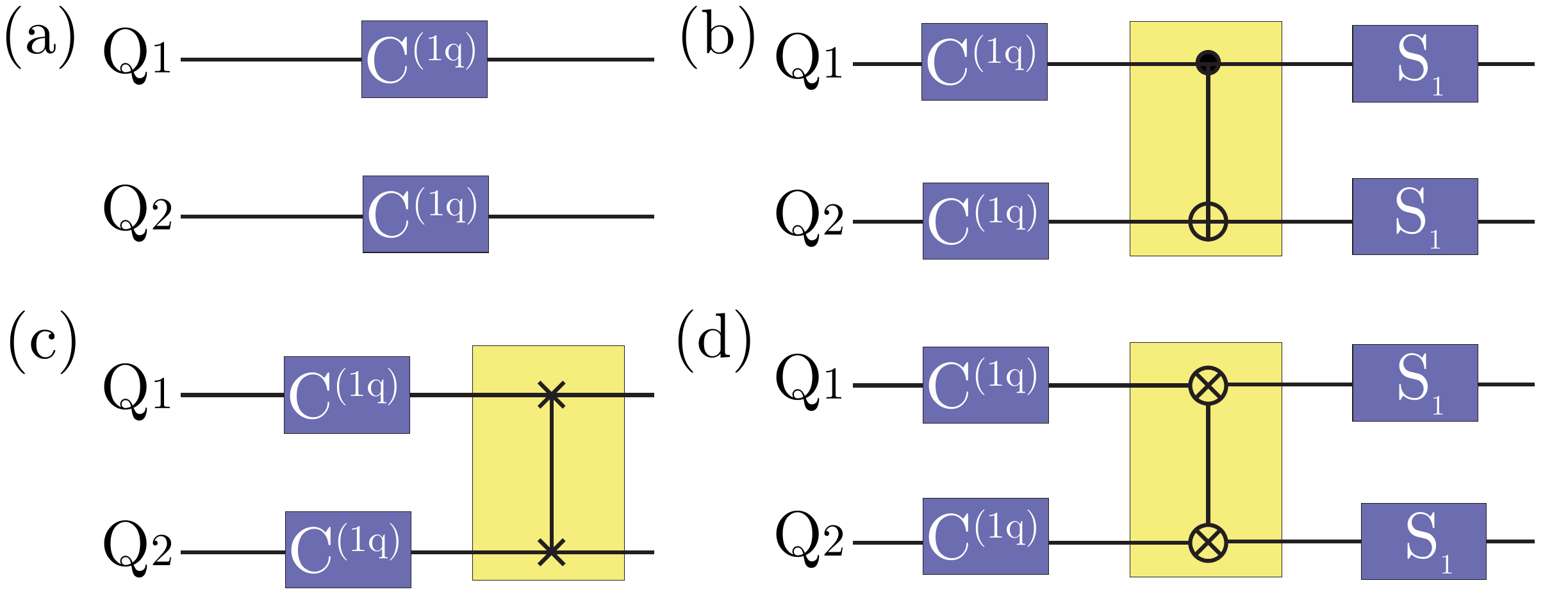}
    \caption{Four classes of Clifford gates used in two-qubit RB: (a) $C_1 \otimes C_1$, (b) CNOT class, (c) SWAP class and (d) iSWAP class. This decomposition results in the minimum number of iSWAP gates when decomposing CNOT and SWAP gates in terms of iSWAP~\citep{corcoles2013process}. Since single-qubit and SWAP operations are non-entangling, these classes of gates require rotations only before the gate.}
    \label{fig:2QCliffords}
\end{figure}
The randomized benchmarking experiment applies randomly selected gates from the Clifford group, followed by a final gate that inverts the entire preceding sequence. The n-qubit Clifford group $C_n$ itself is defined as the normalizer of the n-qubit Pauli group $P_n$:
\begin{equation}
    C_n = \{ U | U P_n U^{\dagger} =P_n\}.
\end{equation}
For single qubit RB, the 24 elements of the Clifford group $C_1$ are generated by the Hadamard (H) and Phase (S) gates
\begin{align}
    H &= \frac{1}{\sqrt{2}}\begin{pmatrix}
        1 && 1 \\
        1 && -1\\
    \end{pmatrix}, \quad
    S = \begin{pmatrix}
        1 && 0 \\
        0 && i\\
    \end{pmatrix}.
\end{align}
\par
\begin{figure}[h!]
    \centering
    \includegraphics[width=\linewidth]{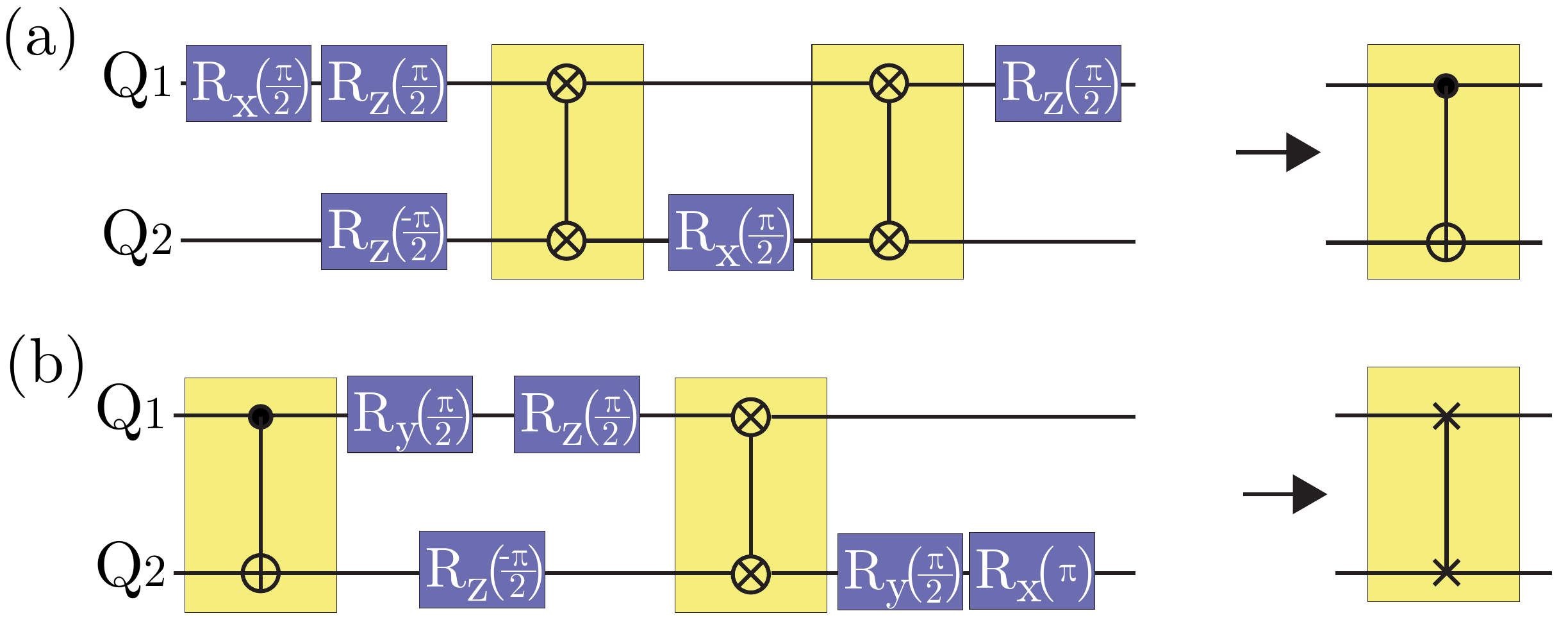}
    \caption{Decomposition of (a) CNOT and (b) SWAP gates in terms of iSWAP gates and single-qubit rotations. Along with single-qubit rotations, implementation of the CNOT gate requires two iSWAPs, while implementation of the SWAP gate requires three iSWAPs.}
    \label{fig:Gate_Decomposition}%
\end{figure}

The two-qubit Clifford group $C_2$ contains 11,520 gates and it is generated by the single-qubit $H$ and $S$ gates as well as the two-qubit CNOT gate. Following ref.~\citep{corcoles2013process}, we efficiently write a generic element of the Clifford group using the minimal number of iSWAP gates by decomposing it according to the following classes: 
\begin{itemize}
    \item The first class, shown in Fig. \ref{fig:2QCliffords}(a), contains 576 ($24^2$) gates, which are formed by the tensor product of single-qubit Clifford gates $C_{1} \otimes C_{1}$.
    \item The second class, also known as the CNOT-like class, contains 5184 gates, which are formed as shown in Fig.~\ref{fig:2QCliffords}(b).
    \item The third class, also known as the SWAP-like class, contains 576 gates, which are formed as shown in Fig.~\ref{fig:2QCliffords}(c). Since SWAP gates are non-entangling, no rotations after the two-qubit gate are needed.
    \item The fourth class, also known as iSWAP-like class, contains 5184 gates which are formed as shown in Fig.~\ref{fig:2QCliffords}(d).
\end{itemize}
In Fig.~\ref{fig:2QCliffords}, $S_1=\{I,S,S^2\}$ where ${S=\exp(-i\pi(X+Y+Z)/3\sqrt{3})}$. In our implementation we further decomposed the CNOT and SWAP gates as the product of single-qubit rotations and iSWAP gates as shown in Fig.~\ref{fig:Gate_Decomposition}.
%
\clearpage
\nocite{*}

\bibliography{reference}

\begin{thebibliography}{42}%
\makeatletter
\providecommand \@ifxundefined [1]{%
 \@ifx{#1\undefined}
}%
\providecommand \@ifnum [1]{%
 \ifnum #1\expandafter \@firstoftwo
 \else \expandafter \@secondoftwo
 \fi
}%
\providecommand \@ifx [1]{%
 \ifx #1\expandafter \@firstoftwo
 \else \expandafter \@secondoftwo
 \fi
}%
\providecommand \natexlab [1]{#1}%
\providecommand \enquote  [1]{``#1''}%
\providecommand \bibnamefont  [1]{#1}%
\providecommand \bibfnamefont [1]{#1}%
\providecommand \citenamefont [1]{#1}%
\providecommand \href@noop [0]{\@secondoftwo}%
\providecommand \href [0]{\begingroup \@sanitize@url \@href}%
\providecommand \@href[1]{\@@startlink{#1}\@@href}%
\providecommand \@@href[1]{\endgroup#1\@@endlink}%
\providecommand \@sanitize@url [0]{\catcode `\\12\catcode `\$12\catcode `\&12\catcode `\#12\catcode `\^12\catcode `\_12\catcode `\%12\relax}%
\providecommand \@@startlink[1]{}%
\providecommand \@@endlink[0]{}%
\providecommand \url  [0]{\begingroup\@sanitize@url \@url }%
\providecommand \@url [1]{\endgroup\@href {#1}{\urlprefix }}%
\providecommand \urlprefix  [0]{URL }%
\providecommand \Eprint [0]{\href }%
\providecommand \doibase [0]{https://doi.org/}%
\providecommand \selectlanguage [0]{\@gobble}%
\providecommand \bibinfo  [0]{\@secondoftwo}%
\providecommand \bibfield  [0]{\@secondoftwo}%
\providecommand \translation [1]{[#1]}%
\providecommand \BibitemOpen [0]{}%
\providecommand \bibitemStop [0]{}%
\providecommand \bibitemNoStop [0]{.\EOS\space}%
\providecommand \EOS [0]{\spacefactor3000\relax}%
\providecommand \BibitemShut  [1]{\csname bibitem#1\endcsname}%
\let\auto@bib@innerbib\@empty
\bibitem [{\citenamefont {Wu}\ \emph {et~al.}(2024)\citenamefont {Wu}, \citenamefont {Yan}, \citenamefont {Andersson}, \citenamefont {Anferov}, \citenamefont {Chou}, \citenamefont {Conner}, \citenamefont {Grebel}, \citenamefont {Joshi}, \citenamefont {Li}, \citenamefont {Miller} \emph {et~al.}}]{wu2024modular}%
  \BibitemOpen
  \bibfield  {author} {\bibinfo {author} {\bibfnamefont {X.}~\bibnamefont {Wu}}, \bibinfo {author} {\bibfnamefont {H.}~\bibnamefont {Yan}}, \bibinfo {author} {\bibfnamefont {G.}~\bibnamefont {Andersson}}, \bibinfo {author} {\bibfnamefont {A.}~\bibnamefont {Anferov}}, \bibinfo {author} {\bibfnamefont {M.-H.}\ \bibnamefont {Chou}}, \bibinfo {author} {\bibfnamefont {C.~R.}\ \bibnamefont {Conner}}, \bibinfo {author} {\bibfnamefont {J.}~\bibnamefont {Grebel}}, \bibinfo {author} {\bibfnamefont {Y.~J.}\ \bibnamefont {Joshi}}, \bibinfo {author} {\bibfnamefont {S.}~\bibnamefont {Li}}, \bibinfo {author} {\bibfnamefont {J.~M.}\ \bibnamefont {Miller}}, \emph {et~al.},\ }\bibfield  {title} {\bibinfo {title} {Modular quantum processor with an all-to-all reconfigurable router},\ }\href {https://doi.org/https://doi.org/10.1103/PhysRevX.14.041030} {\bibfield  {journal} {\bibinfo  {journal} {Physical Review X}\ }\textbf {\bibinfo {volume} {14}},\ \bibinfo {pages} {041030} (\bibinfo {year} {2024})}\BibitemShut {NoStop}%
\bibitem [{\citenamefont {Vall{\'e}s-Sanclemente}\ \emph {et~al.}(2025)\citenamefont {Vall{\'e}s-Sanclemente}, \citenamefont {Vroomans}, \citenamefont {van Abswoude}, \citenamefont {Brulleman}, \citenamefont {Stavenga}, \citenamefont {van~der Meer}, \citenamefont {Xin}, \citenamefont {Lawrence}, \citenamefont {Singh}, \citenamefont {Rol} \emph {et~al.}}]{valles2025optimizing}%
  \BibitemOpen
  \bibfield  {author} {\bibinfo {author} {\bibfnamefont {S.}~\bibnamefont {Vall{\'e}s-Sanclemente}}, \bibinfo {author} {\bibfnamefont {T.}~\bibnamefont {Vroomans}}, \bibinfo {author} {\bibfnamefont {T.}~\bibnamefont {van Abswoude}}, \bibinfo {author} {\bibfnamefont {F.}~\bibnamefont {Brulleman}}, \bibinfo {author} {\bibfnamefont {T.}~\bibnamefont {Stavenga}}, \bibinfo {author} {\bibfnamefont {S.}~\bibnamefont {van~der Meer}}, \bibinfo {author} {\bibfnamefont {Y.}~\bibnamefont {Xin}}, \bibinfo {author} {\bibfnamefont {A.}~\bibnamefont {Lawrence}}, \bibinfo {author} {\bibfnamefont {V.}~\bibnamefont {Singh}}, \bibinfo {author} {\bibfnamefont {M.}~\bibnamefont {Rol}}, \emph {et~al.},\ }\bibfield  {title} {\bibinfo {title} {Optimizing the frequency positioning of tunable couplers in a circuit qed processor to mitigate spectator effects on quantum operations},\ }\bibfield  {journal} {\bibinfo  {journal} {arXiv preprint arXiv:2503.13225}\ }\href {https://doi.org/https://doi.org/10.48550/arXiv.2503.13225}
  {https://doi.org/10.48550/arXiv.2503.13225} (\bibinfo {year} {2025})\BibitemShut {NoStop}%
\bibitem [{\citenamefont {Li}\ \emph {et~al.}(2024)\citenamefont {Li}, \citenamefont {Kubo}, \citenamefont {Ho}, \citenamefont {Yan}, \citenamefont {Nakamura},\ and\ \citenamefont {Goto}}]{li2024realization}%
  \BibitemOpen
  \bibfield  {author} {\bibinfo {author} {\bibfnamefont {R.}~\bibnamefont {Li}}, \bibinfo {author} {\bibfnamefont {K.}~\bibnamefont {Kubo}}, \bibinfo {author} {\bibfnamefont {Y.}~\bibnamefont {Ho}}, \bibinfo {author} {\bibfnamefont {Z.}~\bibnamefont {Yan}}, \bibinfo {author} {\bibfnamefont {Y.}~\bibnamefont {Nakamura}},\ and\ \bibinfo {author} {\bibfnamefont {H.}~\bibnamefont {Goto}},\ }\bibfield  {title} {\bibinfo {title} {Realization of high-fidelity cz gate based on a double-transmon coupler},\ }\href {https://doi.org/10.1103/PhysRevX.14.041050} {\bibfield  {journal} {\bibinfo  {journal} {Physical Review X}\ }\textbf {\bibinfo {volume} {14}},\ \bibinfo {pages} {041050} (\bibinfo {year} {2024})}\BibitemShut {NoStop}%
\bibitem [{\citenamefont {Kim}\ \emph {et~al.}(2026)\citenamefont {Kim}, \citenamefont {Butler},\ and\ \citenamefont {Painter}}]{kim2026tunable}%
  \BibitemOpen
  \bibfield  {author} {\bibinfo {author} {\bibfnamefont {G.}~\bibnamefont {Kim}}, \bibinfo {author} {\bibfnamefont {A.}~\bibnamefont {Butler}},\ and\ \bibinfo {author} {\bibfnamefont {O.}~\bibnamefont {Painter}},\ }\bibfield  {title} {\bibinfo {title} {A tunable, modeless, and hybridization-free cross-kerr coupler for miniaturized superconducting qubits},\ }\bibfield  {journal} {\bibinfo  {journal} {arXiv preprint arXiv:2602.03186}\ }\href {https://doi.org/https://doi.org/10.48550/arXiv.2602.03186} {https://doi.org/10.48550/arXiv.2602.03186} (\bibinfo {year} {2026})\BibitemShut {NoStop}%
\bibitem [{\citenamefont {Koshino}(2025)}]{koshino2025galvanically}%
  \BibitemOpen
  \bibfield  {author} {\bibinfo {author} {\bibfnamefont {K.}~\bibnamefont {Koshino}},\ }\bibfield  {title} {\bibinfo {title} {Galvanically connected tunable coupler between a cavity and a waveguide},\ }\href {https://doi.org/10.1088/1367-2630/adc1fb} {\bibfield  {journal} {\bibinfo  {journal} {New Journal of Physics}\ }\textbf {\bibinfo {volume} {27}},\ \bibinfo {pages} {043001} (\bibinfo {year} {2025})}\BibitemShut {NoStop}%
\bibitem [{\citenamefont {Xu}\ \emph {et~al.}(2026)\citenamefont {Xu}, \citenamefont {Deng}, \citenamefont {Zheng}, \citenamefont {Yan}, \citenamefont {Zhang}, \citenamefont {Zhang}, \citenamefont {Huang}, \citenamefont {Xia}, \citenamefont {Liao}, \citenamefont {Zhang} \emph {et~al.}}]{xu2026tunable}%
  \BibitemOpen
  \bibfield  {author} {\bibinfo {author} {\bibfnamefont {J.}~\bibnamefont {Xu}}, \bibinfo {author} {\bibfnamefont {X.}~\bibnamefont {Deng}}, \bibinfo {author} {\bibfnamefont {W.}~\bibnamefont {Zheng}}, \bibinfo {author} {\bibfnamefont {W.}~\bibnamefont {Yan}}, \bibinfo {author} {\bibfnamefont {T.}~\bibnamefont {Zhang}}, \bibinfo {author} {\bibfnamefont {Z.}~\bibnamefont {Zhang}}, \bibinfo {author} {\bibfnamefont {W.}~\bibnamefont {Huang}}, \bibinfo {author} {\bibfnamefont {X.}~\bibnamefont {Xia}}, \bibinfo {author} {\bibfnamefont {X.}~\bibnamefont {Liao}}, \bibinfo {author} {\bibfnamefont {Y.}~\bibnamefont {Zhang}}, \emph {et~al.},\ }\bibfield  {title} {\bibinfo {title} {Tunable hybrid-mode coupler enabling strong interactions between transmons at centimeter-scale distance},\ }\href {https://doi.org/https://doi.org/10.1103/ls5b-279m} {\bibfield  {journal} {\bibinfo  {journal} {Physical Review Applied}\ }\textbf {\bibinfo {volume} {25}},\ \bibinfo {pages} {014016} (\bibinfo {year} {2026})}\BibitemShut
  {NoStop}%
\bibitem [{\citenamefont {Zhang}\ \emph {et~al.}(2024)\citenamefont {Zhang}, \citenamefont {Ding}, \citenamefont {Weiss}, \citenamefont {Huang}, \citenamefont {Ma}, \citenamefont {Guinn}, \citenamefont {Sussman}, \citenamefont {Chitta}, \citenamefont {Chen}, \citenamefont {Houck} \emph {et~al.}}]{zhang2024tunable}%
  \BibitemOpen
  \bibfield  {author} {\bibinfo {author} {\bibfnamefont {H.}~\bibnamefont {Zhang}}, \bibinfo {author} {\bibfnamefont {C.}~\bibnamefont {Ding}}, \bibinfo {author} {\bibfnamefont {D.}~\bibnamefont {Weiss}}, \bibinfo {author} {\bibfnamefont {Z.}~\bibnamefont {Huang}}, \bibinfo {author} {\bibfnamefont {Y.}~\bibnamefont {Ma}}, \bibinfo {author} {\bibfnamefont {C.}~\bibnamefont {Guinn}}, \bibinfo {author} {\bibfnamefont {S.}~\bibnamefont {Sussman}}, \bibinfo {author} {\bibfnamefont {S.~P.}\ \bibnamefont {Chitta}}, \bibinfo {author} {\bibfnamefont {D.}~\bibnamefont {Chen}}, \bibinfo {author} {\bibfnamefont {A.~A.}\ \bibnamefont {Houck}}, \emph {et~al.},\ }\bibfield  {title} {\bibinfo {title} {Tunable inductive coupler for high-fidelity gates between fluxonium qubits},\ }\href {https://doi.org/https://doi.org/10.1103/PRXQuantum.5.020326} {\bibfield  {journal} {\bibinfo  {journal} {PRX Quantum}\ }\textbf {\bibinfo {volume} {5}},\ \bibinfo {pages} {020326} (\bibinfo {year} {2024})}\BibitemShut {NoStop}%
\bibitem [{\citenamefont {Jin}\ \emph {et~al.}(2025)\citenamefont {Jin}, \citenamefont {Parrott}, \citenamefont {Cicak}, \citenamefont {Kotler}, \citenamefont {Lecocq}, \citenamefont {Teufel}, \citenamefont {Aumentado}, \citenamefont {Kapit},\ and\ \citenamefont {Simmonds}}]{jin2025superconducting}%
  \BibitemOpen
  \bibfield  {author} {\bibinfo {author} {\bibfnamefont {X.}~\bibnamefont {Jin}}, \bibinfo {author} {\bibfnamefont {Z.}~\bibnamefont {Parrott}}, \bibinfo {author} {\bibfnamefont {K.}~\bibnamefont {Cicak}}, \bibinfo {author} {\bibfnamefont {S.}~\bibnamefont {Kotler}}, \bibinfo {author} {\bibfnamefont {F.}~\bibnamefont {Lecocq}}, \bibinfo {author} {\bibfnamefont {J.}~\bibnamefont {Teufel}}, \bibinfo {author} {\bibfnamefont {J.}~\bibnamefont {Aumentado}}, \bibinfo {author} {\bibfnamefont {E.}~\bibnamefont {Kapit}},\ and\ \bibinfo {author} {\bibfnamefont {R.}~\bibnamefont {Simmonds}},\ }\bibfield  {title} {\bibinfo {title} {Superconducting architecture demonstrating fast, tunable high-fidelity cz gates with parametric control of zz coupling},\ }\href {https://doi.org/https://doi.org/10.1103/kmls-lgp5} {\bibfield  {journal} {\bibinfo  {journal} {Physical Review Applied}\ }\textbf {\bibinfo {volume} {24}},\ \bibinfo {pages} {064026} (\bibinfo {year} {2025})}\BibitemShut {NoStop}%
\bibitem [{\citenamefont {Chakraborty}\ \emph {et~al.}(2025)\citenamefont {Chakraborty}, \citenamefont {Bhandari}, \citenamefont {Brise{\~n}o-Colunga}, \citenamefont {Stevenson}, \citenamefont {Pedramrazi}, \citenamefont {Liu}, \citenamefont {Santiago}, \citenamefont {Siddiqi}, \citenamefont {Dressel},\ and\ \citenamefont {Jordan}}]{chakraborty2025tunable}%
  \BibitemOpen
  \bibfield  {author} {\bibinfo {author} {\bibfnamefont {A.}~\bibnamefont {Chakraborty}}, \bibinfo {author} {\bibfnamefont {B.}~\bibnamefont {Bhandari}}, \bibinfo {author} {\bibfnamefont {D.~D.}\ \bibnamefont {Brise{\~n}o-Colunga}}, \bibinfo {author} {\bibfnamefont {N.}~\bibnamefont {Stevenson}}, \bibinfo {author} {\bibfnamefont {Z.}~\bibnamefont {Pedramrazi}}, \bibinfo {author} {\bibfnamefont {C.-H.}\ \bibnamefont {Liu}}, \bibinfo {author} {\bibfnamefont {D.~I.}\ \bibnamefont {Santiago}}, \bibinfo {author} {\bibfnamefont {I.}~\bibnamefont {Siddiqi}}, \bibinfo {author} {\bibfnamefont {J.}~\bibnamefont {Dressel}},\ and\ \bibinfo {author} {\bibfnamefont {A.~N.}\ \bibnamefont {Jordan}},\ }\bibfield  {title} {\bibinfo {title} {Tunable superconducting quantum interference device coupler for fluxonium qubits},\ }\bibfield  {journal} {\bibinfo  {journal} {arXiv preprint arXiv:2508.16907}\ }\href {https://doi.org/https://doi.org/10.48550/arXiv.2508.16907} {https://doi.org/10.48550/arXiv.2508.16907} (\bibinfo {year}
  {2025})\BibitemShut {NoStop}%
\bibitem [{\citenamefont {Subramanian}\ and\ \citenamefont {Lupascu}(2023)}]{subramanian2023efficient}%
  \BibitemOpen
  \bibfield  {author} {\bibinfo {author} {\bibfnamefont {M.}~\bibnamefont {Subramanian}}\ and\ \bibinfo {author} {\bibfnamefont {A.}~\bibnamefont {Lupascu}},\ }\bibfield  {title} {\bibinfo {title} {Efficient two-qutrit gates in superconducting circuits using parametric coupling},\ }\href {https://doi.org/https://doi.org/10.1103/PhysRevA.108.062616} {\bibfield  {journal} {\bibinfo  {journal} {Physical Review A}\ }\textbf {\bibinfo {volume} {108}},\ \bibinfo {pages} {062616} (\bibinfo {year} {2023})}\BibitemShut {NoStop}%
\bibitem [{\citenamefont {Warren}\ \emph {et~al.}(2023)\citenamefont {Warren}, \citenamefont {Fern{\'a}ndez-Pend{\'a}s}, \citenamefont {Ahmed}, \citenamefont {Abad}, \citenamefont {Bengtsson}, \citenamefont {Bizn{\'a}rov{\'a}}, \citenamefont {Debnath}, \citenamefont {Gu}, \citenamefont {Kri{\v{z}}an}, \citenamefont {Osman} \emph {et~al.}}]{warren2023extensive}%
  \BibitemOpen
  \bibfield  {author} {\bibinfo {author} {\bibfnamefont {C.~W.}\ \bibnamefont {Warren}}, \bibinfo {author} {\bibfnamefont {J.}~\bibnamefont {Fern{\'a}ndez-Pend{\'a}s}}, \bibinfo {author} {\bibfnamefont {S.}~\bibnamefont {Ahmed}}, \bibinfo {author} {\bibfnamefont {T.}~\bibnamefont {Abad}}, \bibinfo {author} {\bibfnamefont {A.}~\bibnamefont {Bengtsson}}, \bibinfo {author} {\bibfnamefont {J.}~\bibnamefont {Bizn{\'a}rov{\'a}}}, \bibinfo {author} {\bibfnamefont {K.}~\bibnamefont {Debnath}}, \bibinfo {author} {\bibfnamefont {X.}~\bibnamefont {Gu}}, \bibinfo {author} {\bibfnamefont {C.}~\bibnamefont {Kri{\v{z}}an}}, \bibinfo {author} {\bibfnamefont {A.}~\bibnamefont {Osman}}, \emph {et~al.},\ }\bibfield  {title} {\bibinfo {title} {Extensive characterization and implementation of a family of three-qubit gates at the coherence limit},\ }\href {https://doi.org/https://doi.org/10.1038/s41534-023-00711-x} {\bibfield  {journal} {\bibinfo  {journal} {npj Quantum Information}\ }\textbf {\bibinfo {volume} {9}},\ \bibinfo
  {pages} {44} (\bibinfo {year} {2023})}\BibitemShut {NoStop}%
\bibitem [{\citenamefont {Sung}\ \emph {et~al.}(2021)\citenamefont {Sung}, \citenamefont {Ding}, \citenamefont {Braum{\"u}ller}, \citenamefont {Veps{\"a}l{\"a}inen}, \citenamefont {Kannan}, \citenamefont {Kjaergaard}, \citenamefont {Greene}, \citenamefont {Samach}, \citenamefont {McNally}, \citenamefont {Kim} \emph {et~al.}}]{sung2021realization}%
  \BibitemOpen
  \bibfield  {author} {\bibinfo {author} {\bibfnamefont {Y.}~\bibnamefont {Sung}}, \bibinfo {author} {\bibfnamefont {L.}~\bibnamefont {Ding}}, \bibinfo {author} {\bibfnamefont {J.}~\bibnamefont {Braum{\"u}ller}}, \bibinfo {author} {\bibfnamefont {A.}~\bibnamefont {Veps{\"a}l{\"a}inen}}, \bibinfo {author} {\bibfnamefont {B.}~\bibnamefont {Kannan}}, \bibinfo {author} {\bibfnamefont {M.}~\bibnamefont {Kjaergaard}}, \bibinfo {author} {\bibfnamefont {A.}~\bibnamefont {Greene}}, \bibinfo {author} {\bibfnamefont {G.~O.}\ \bibnamefont {Samach}}, \bibinfo {author} {\bibfnamefont {C.}~\bibnamefont {McNally}}, \bibinfo {author} {\bibfnamefont {D.}~\bibnamefont {Kim}}, \emph {et~al.},\ }\bibfield  {title} {\bibinfo {title} {Realization of high-fidelity cz and zz-free iswap gates with a tunable coupler},\ }\href {https://doi.org/https://doi.org/10.1103/PhysRevX.11.021058} {\bibfield  {journal} {\bibinfo  {journal} {Physical Review X}\ }\textbf {\bibinfo {volume} {11}},\ \bibinfo {pages} {021058} (\bibinfo {year}
  {2021})}\BibitemShut {NoStop}%
\bibitem [{\citenamefont {Abrams}\ \emph {et~al.}(2020)\citenamefont {Abrams}, \citenamefont {Didier}, \citenamefont {Johnson}, \citenamefont {Silva},\ and\ \citenamefont {Ryan}}]{Abrams2020}%
  \BibitemOpen
  \bibfield  {author} {\bibinfo {author} {\bibfnamefont {D.~M.}\ \bibnamefont {Abrams}}, \bibinfo {author} {\bibfnamefont {N.}~\bibnamefont {Didier}}, \bibinfo {author} {\bibfnamefont {B.~R.}\ \bibnamefont {Johnson}}, \bibinfo {author} {\bibfnamefont {M.~P.~d.}\ \bibnamefont {Silva}},\ and\ \bibinfo {author} {\bibfnamefont {C.~A.}\ \bibnamefont {Ryan}},\ }\bibfield  {title} {\bibinfo {title} {Implementation of xy entangling gates with a single calibrated pulse},\ }\href {https://doi.org/10.1038/s41928-020-00498-1} {\bibfield  {journal} {\bibinfo  {journal} {Nature Electronics}\ }\textbf {\bibinfo {volume} {3}},\ \bibinfo {pages} {744} (\bibinfo {year} {2020})}\BibitemShut {NoStop}%
\bibitem [{\citenamefont {Ganzhorn}\ \emph {et~al.}(2019)\citenamefont {Ganzhorn}, \citenamefont {Egger}, \citenamefont {Barkoutsos}, \citenamefont {Ollitrault}, \citenamefont {Salis}, \citenamefont {Moll}, \citenamefont {Roth}, \citenamefont {Fuhrer}, \citenamefont {Mueller}, \citenamefont {Woerner}, \citenamefont {Tavernelli},\ and\ \citenamefont {Filipp}}]{Ganzhorn2019}%
  \BibitemOpen
  \bibfield  {author} {\bibinfo {author} {\bibfnamefont {M.}~\bibnamefont {Ganzhorn}}, \bibinfo {author} {\bibfnamefont {D.}~\bibnamefont {Egger}}, \bibinfo {author} {\bibfnamefont {P.}~\bibnamefont {Barkoutsos}}, \bibinfo {author} {\bibfnamefont {P.}~\bibnamefont {Ollitrault}}, \bibinfo {author} {\bibfnamefont {G.}~\bibnamefont {Salis}}, \bibinfo {author} {\bibfnamefont {N.}~\bibnamefont {Moll}}, \bibinfo {author} {\bibfnamefont {M.}~\bibnamefont {Roth}}, \bibinfo {author} {\bibfnamefont {A.}~\bibnamefont {Fuhrer}}, \bibinfo {author} {\bibfnamefont {P.}~\bibnamefont {Mueller}}, \bibinfo {author} {\bibfnamefont {S.}~\bibnamefont {Woerner}}, \bibinfo {author} {\bibfnamefont {I.}~\bibnamefont {Tavernelli}},\ and\ \bibinfo {author} {\bibfnamefont {S.}~\bibnamefont {Filipp}},\ }\bibfield  {title} {\bibinfo {title} {Gate-efficient simulation of molecular eigenstates on a quantum computer},\ }\href {https://doi.org/10.1103/PhysRevApplied.11.044092} {\bibfield  {journal} {\bibinfo  {journal} {Phys. Rev. Appl.}\
  }\textbf {\bibinfo {volume} {11}},\ \bibinfo {pages} {044092} (\bibinfo {year} {2019})}\BibitemShut {NoStop}%
\bibitem [{\citenamefont {McEwen}\ \emph {et~al.}(2023)\citenamefont {McEwen}, \citenamefont {Bacon},\ and\ \citenamefont {Gidney}}]{mcewen2023relaxing}%
  \BibitemOpen
  \bibfield  {author} {\bibinfo {author} {\bibfnamefont {M.}~\bibnamefont {McEwen}}, \bibinfo {author} {\bibfnamefont {D.}~\bibnamefont {Bacon}},\ and\ \bibinfo {author} {\bibfnamefont {C.}~\bibnamefont {Gidney}},\ }\bibfield  {title} {\bibinfo {title} {Relaxing hardware requirements for surface code circuits using time-dynamics},\ }\href {https://doi.org/https://doi.org/10.22331/q-2023-11-07-1172} {\bibfield  {journal} {\bibinfo  {journal} {Quantum}\ }\textbf {\bibinfo {volume} {7}},\ \bibinfo {pages} {1172} (\bibinfo {year} {2023})}\BibitemShut {NoStop}%
\bibitem [{\citenamefont {Campbell}\ \emph {et~al.}(2023)\citenamefont {Campbell}, \citenamefont {Kamal}, \citenamefont {Ranzani}, \citenamefont {Senatore},\ and\ \citenamefont {LaHaye}}]{campbell2023modular}%
  \BibitemOpen
  \bibfield  {author} {\bibinfo {author} {\bibfnamefont {D.~L.}\ \bibnamefont {Campbell}}, \bibinfo {author} {\bibfnamefont {A.}~\bibnamefont {Kamal}}, \bibinfo {author} {\bibfnamefont {L.}~\bibnamefont {Ranzani}}, \bibinfo {author} {\bibfnamefont {M.}~\bibnamefont {Senatore}},\ and\ \bibinfo {author} {\bibfnamefont {M.~D.}\ \bibnamefont {LaHaye}},\ }\bibfield  {title} {\bibinfo {title} {Modular tunable coupler for superconducting circuits},\ }\href {https://doi.org/https://doi.org/10.1103/PhysRevApplied.19.064043} {\bibfield  {journal} {\bibinfo  {journal} {Physical Review Applied}\ }\textbf {\bibinfo {volume} {19}},\ \bibinfo {pages} {064043} (\bibinfo {year} {2023})}\BibitemShut {NoStop}%
\bibitem [{\citenamefont {Goto}(2022)}]{goto2022double}%
  \BibitemOpen
  \bibfield  {author} {\bibinfo {author} {\bibfnamefont {H.}~\bibnamefont {Goto}},\ }\bibfield  {title} {\bibinfo {title} {Double-transmon coupler: Fast two-qubit gate with no residual coupling for highly detuned superconducting qubits},\ }\href {https://doi.org/https://doi.org/10.1103/PhysRevApplied.18.034038} {\bibfield  {journal} {\bibinfo  {journal} {Physical review applied}\ }\textbf {\bibinfo {volume} {18}},\ \bibinfo {pages} {034038} (\bibinfo {year} {2022})}\BibitemShut {NoStop}%
\bibitem [{\citenamefont {Campbell}\ \emph {et~al.}(2026)\citenamefont {Campbell}, \citenamefont {McCoy}, \citenamefont {Andrews}, \citenamefont {Madden}, \citenamefont {Horowitz}, \citenamefont {Husremovi{\'c}}, \citenamefont {Marash}, \citenamefont {Nadeau}, \citenamefont {Nguyen}, \citenamefont {Senatore} \emph {et~al.}}]{campbell2026transmon}%
  \BibitemOpen
  \bibfield  {author} {\bibinfo {author} {\bibfnamefont {D.~L.}\ \bibnamefont {Campbell}}, \bibinfo {author} {\bibfnamefont {S.}~\bibnamefont {McCoy}}, \bibinfo {author} {\bibfnamefont {M.}~\bibnamefont {Andrews}}, \bibinfo {author} {\bibfnamefont {A.}~\bibnamefont {Madden}}, \bibinfo {author} {\bibfnamefont {V.~R.}\ \bibnamefont {Horowitz}}, \bibinfo {author} {\bibfnamefont {B.}~\bibnamefont {Husremovi{\'c}}}, \bibinfo {author} {\bibfnamefont {S.}~\bibnamefont {Marash}}, \bibinfo {author} {\bibfnamefont {C.}~\bibnamefont {Nadeau}}, \bibinfo {author} {\bibfnamefont {M.}~\bibnamefont {Nguyen}}, \bibinfo {author} {\bibfnamefont {M.}~\bibnamefont {Senatore}}, \emph {et~al.},\ }\bibfield  {title} {\bibinfo {title} {Transmon architecture for emission and detection of single microwave photons},\ }\bibfield  {journal} {\bibinfo  {journal} {arXiv preprint arXiv:2601.11378}\ }\href {https://doi.org/https://doi.org/10.48550/arXiv.2601.11378} {https://doi.org/10.48550/arXiv.2601.11378} (\bibinfo {year}
  {2026})\BibitemShut {NoStop}%
\bibitem [{\citenamefont {Heunisch}\ \emph {et~al.}(2023)\citenamefont {Heunisch}, \citenamefont {Eichler},\ and\ \citenamefont {Hartmann}}]{heunisch2023tunable}%
  \BibitemOpen
  \bibfield  {author} {\bibinfo {author} {\bibfnamefont {L.}~\bibnamefont {Heunisch}}, \bibinfo {author} {\bibfnamefont {C.}~\bibnamefont {Eichler}},\ and\ \bibinfo {author} {\bibfnamefont {M.~J.}\ \bibnamefont {Hartmann}},\ }\bibfield  {title} {\bibinfo {title} {Tunable coupler to fully decouple and maximally localize superconducting qubits},\ }\href {https://doi.org/https://doi.org/10.1103/PhysRevApplied.20.064037} {\bibfield  {journal} {\bibinfo  {journal} {Physical Review Applied}\ }\textbf {\bibinfo {volume} {20}},\ \bibinfo {pages} {064037} (\bibinfo {year} {2023})}\BibitemShut {NoStop}%
\bibitem [{\citenamefont {Li}\ \emph {et~al.}(2025)\citenamefont {Li}, \citenamefont {Kubo}, \citenamefont {Ho}, \citenamefont {Yan}, \citenamefont {Inoue}, \citenamefont {Nakamura},\ and\ \citenamefont {Goto}}]{li2025capacitively}%
  \BibitemOpen
  \bibfield  {author} {\bibinfo {author} {\bibfnamefont {R.}~\bibnamefont {Li}}, \bibinfo {author} {\bibfnamefont {K.}~\bibnamefont {Kubo}}, \bibinfo {author} {\bibfnamefont {Y.}~\bibnamefont {Ho}}, \bibinfo {author} {\bibfnamefont {Z.}~\bibnamefont {Yan}}, \bibinfo {author} {\bibfnamefont {S.}~\bibnamefont {Inoue}}, \bibinfo {author} {\bibfnamefont {Y.}~\bibnamefont {Nakamura}},\ and\ \bibinfo {author} {\bibfnamefont {H.}~\bibnamefont {Goto}},\ }\bibfield  {title} {\bibinfo {title} {Capacitively shunted double-transmon coupler realizing bias-free idling and a high-fidelity cz gate},\ }\href {https://doi.org/https://doi.org/10.1103/l8tq-7sb3} {\bibfield  {journal} {\bibinfo  {journal} {Physical Review Applied}\ }\textbf {\bibinfo {volume} {23}},\ \bibinfo {pages} {064069} (\bibinfo {year} {2025})}\BibitemShut {NoStop}%
\bibitem [{\citenamefont {Yuan}\ \emph {et~al.}(2026)\citenamefont {Yuan}, \citenamefont {Zhang}, \citenamefont {He}, \citenamefont {Fei}, \citenamefont {Han}, \citenamefont {Wang}, \citenamefont {Sun}, \citenamefont {Mu}, \citenamefont {Zhao}, \citenamefont {Liu} \emph {et~al.}}]{yuan2026tunable}%
  \BibitemOpen
  \bibfield  {author} {\bibinfo {author} {\bibfnamefont {B.}~\bibnamefont {Yuan}}, \bibinfo {author} {\bibfnamefont {C.}~\bibnamefont {Zhang}}, \bibinfo {author} {\bibfnamefont {H.}~\bibnamefont {He}}, \bibinfo {author} {\bibfnamefont {Y.}~\bibnamefont {Fei}}, \bibinfo {author} {\bibfnamefont {C.}~\bibnamefont {Han}}, \bibinfo {author} {\bibfnamefont {S.}~\bibnamefont {Wang}}, \bibinfo {author} {\bibfnamefont {H.}~\bibnamefont {Sun}}, \bibinfo {author} {\bibfnamefont {Q.}~\bibnamefont {Mu}}, \bibinfo {author} {\bibfnamefont {B.}~\bibnamefont {Zhao}}, \bibinfo {author} {\bibfnamefont {F.}~\bibnamefont {Liu}}, \emph {et~al.},\ }\bibfield  {title} {\bibinfo {title} {Tunable nonlocal zz interaction for remote controlled-z gates between distributed fixed-frequency qubits},\ }\bibfield  {journal} {\bibinfo  {journal} {arXiv preprint arXiv:2603.28526}\ }\href {https://doi.org/https://doi.org/10.48550/arXiv.2603.28526} {https://doi.org/10.48550/arXiv.2603.28526} (\bibinfo {year} {2026})\BibitemShut {NoStop}%
\bibitem [{\citenamefont {Kimmel}\ \emph {et~al.}(2015)\citenamefont {Kimmel}, \citenamefont {Low},\ and\ \citenamefont {Yoder}}]{kimmel2015robust}%
  \BibitemOpen
  \bibfield  {author} {\bibinfo {author} {\bibfnamefont {S.}~\bibnamefont {Kimmel}}, \bibinfo {author} {\bibfnamefont {G.~H.}\ \bibnamefont {Low}},\ and\ \bibinfo {author} {\bibfnamefont {T.~J.}\ \bibnamefont {Yoder}},\ }\bibfield  {title} {\bibinfo {title} {Robust calibration of a universal single-qubit gate set via robust phase estimation},\ }\href {https://doi.org/https://doi.org/10.1103/PhysRevA.92.062315} {\bibfield  {journal} {\bibinfo  {journal} {Physical Review A}\ }\textbf {\bibinfo {volume} {92}},\ \bibinfo {pages} {062315} (\bibinfo {year} {2015})}\BibitemShut {NoStop}%
\bibitem [{\citenamefont {Rudinger}\ \emph {et~al.}(2025)\citenamefont {Rudinger}, \citenamefont {Marceaux}, \citenamefont {Hashim}, \citenamefont {Santiago}, \citenamefont {Siddiqi},\ and\ \citenamefont {Young}}]{rudinger2025heisenberg}%
  \BibitemOpen
  \bibfield  {author} {\bibinfo {author} {\bibfnamefont {K.}~\bibnamefont {Rudinger}}, \bibinfo {author} {\bibfnamefont {J.}~\bibnamefont {Marceaux}}, \bibinfo {author} {\bibfnamefont {A.}~\bibnamefont {Hashim}}, \bibinfo {author} {\bibfnamefont {D.~I.}\ \bibnamefont {Santiago}}, \bibinfo {author} {\bibfnamefont {I.}~\bibnamefont {Siddiqi}},\ and\ \bibinfo {author} {\bibfnamefont {K.~C.}\ \bibnamefont {Young}},\ }\bibfield  {title} {\bibinfo {title} {Heisenberg-limited calibration of entangling gates with robust phase estimation},\ }\bibfield  {journal} {\bibinfo  {journal} {arXiv preprint arXiv:2502.06698}\ }\href {https://doi.org/https://doi.org/10.48550/arXiv.2502.06698} {https://doi.org/10.48550/arXiv.2502.06698} (\bibinfo {year} {2025})\BibitemShut {NoStop}%
\bibitem [{\citenamefont {Magesan}\ and\ \citenamefont {Gambetta}(2020)}]{magesan2020effective}%
  \BibitemOpen
  \bibfield  {author} {\bibinfo {author} {\bibfnamefont {E.}~\bibnamefont {Magesan}}\ and\ \bibinfo {author} {\bibfnamefont {J.~M.}\ \bibnamefont {Gambetta}},\ }\bibfield  {title} {\bibinfo {title} {Effective hamiltonian models of the cross-resonance gate},\ }\href {https://doi.org/https://doi.org/10.1103/PhysRevA.101.052308} {\bibfield  {journal} {\bibinfo  {journal} {Physical Review A}\ }\textbf {\bibinfo {volume} {101}},\ \bibinfo {pages} {052308} (\bibinfo {year} {2020})}\BibitemShut {NoStop}%
\bibitem [{\citenamefont {Solgun}\ and\ \citenamefont {Srinivasan}(2022)}]{solgun2022direct}%
  \BibitemOpen
  \bibfield  {author} {\bibinfo {author} {\bibfnamefont {F.}~\bibnamefont {Solgun}}\ and\ \bibinfo {author} {\bibfnamefont {S.}~\bibnamefont {Srinivasan}},\ }\bibfield  {title} {\bibinfo {title} {Direct calculation of zz interaction rates in multimode circuit quantum electrodynamics},\ }\href {https://doi.org/https://doi.org/10.1103/PhysRevApplied.18.044025} {\bibfield  {journal} {\bibinfo  {journal} {Physical Review Applied}\ }\textbf {\bibinfo {volume} {18}},\ \bibinfo {pages} {044025} (\bibinfo {year} {2022})}\BibitemShut {NoStop}%
\bibitem [{\citenamefont {Ku}\ \emph {et~al.}(2020)\citenamefont {Ku}, \citenamefont {Xu}, \citenamefont {Brink}, \citenamefont {McKay}, \citenamefont {Hertzberg}, \citenamefont {Ansari},\ and\ \citenamefont {Plourde}}]{ku2020suppression}%
  \BibitemOpen
  \bibfield  {author} {\bibinfo {author} {\bibfnamefont {J.}~\bibnamefont {Ku}}, \bibinfo {author} {\bibfnamefont {X.}~\bibnamefont {Xu}}, \bibinfo {author} {\bibfnamefont {M.}~\bibnamefont {Brink}}, \bibinfo {author} {\bibfnamefont {D.~C.}\ \bibnamefont {McKay}}, \bibinfo {author} {\bibfnamefont {J.~B.}\ \bibnamefont {Hertzberg}}, \bibinfo {author} {\bibfnamefont {M.~H.}\ \bibnamefont {Ansari}},\ and\ \bibinfo {author} {\bibfnamefont {B.}~\bibnamefont {Plourde}},\ }\bibfield  {title} {\bibinfo {title} {Suppression of unwanted zz interactions in a hybrid two-qubit system},\ }\href {https://doi.org/https://doi.org/10.1103/PhysRevLett.125.200504} {\bibfield  {journal} {\bibinfo  {journal} {Physical review letters}\ }\textbf {\bibinfo {volume} {125}},\ \bibinfo {pages} {200504} (\bibinfo {year} {2020})}\BibitemShut {NoStop}%
\bibitem [{\citenamefont {Groszkowski}\ and\ \citenamefont {Koch}(2021)}]{groszkowski2021scqubits}%
  \BibitemOpen
  \bibfield  {author} {\bibinfo {author} {\bibfnamefont {P.}~\bibnamefont {Groszkowski}}\ and\ \bibinfo {author} {\bibfnamefont {J.}~\bibnamefont {Koch}},\ }\bibfield  {title} {\bibinfo {title} {Scqubits: a python package for superconducting qubits},\ }\href {https://doi.org/https://doi.org/10.22331/q-2021-11-17-583} {\bibfield  {journal} {\bibinfo  {journal} {Quantum}\ }\textbf {\bibinfo {volume} {5}},\ \bibinfo {pages} {583} (\bibinfo {year} {2021})}\BibitemShut {NoStop}%
\bibitem [{\citenamefont {Gambetta}\ \emph {et~al.}(2012)\citenamefont {Gambetta}, \citenamefont {C{\'o}rcoles}, \citenamefont {Merkel}, \citenamefont {Johnson}, \citenamefont {Smolin}, \citenamefont {Chow}, \citenamefont {Ryan}, \citenamefont {Rigetti}, \citenamefont {Poletto}, \citenamefont {Ohki} \emph {et~al.}}]{gambetta2012characterization}%
  \BibitemOpen
  \bibfield  {author} {\bibinfo {author} {\bibfnamefont {J.~M.}\ \bibnamefont {Gambetta}}, \bibinfo {author} {\bibfnamefont {A.~D.}\ \bibnamefont {C{\'o}rcoles}}, \bibinfo {author} {\bibfnamefont {S.~T.}\ \bibnamefont {Merkel}}, \bibinfo {author} {\bibfnamefont {B.~R.}\ \bibnamefont {Johnson}}, \bibinfo {author} {\bibfnamefont {J.~A.}\ \bibnamefont {Smolin}}, \bibinfo {author} {\bibfnamefont {J.~M.}\ \bibnamefont {Chow}}, \bibinfo {author} {\bibfnamefont {C.~A.}\ \bibnamefont {Ryan}}, \bibinfo {author} {\bibfnamefont {C.}~\bibnamefont {Rigetti}}, \bibinfo {author} {\bibfnamefont {S.}~\bibnamefont {Poletto}}, \bibinfo {author} {\bibfnamefont {T.~A.}\ \bibnamefont {Ohki}}, \emph {et~al.},\ }\bibfield  {title} {\bibinfo {title} {Characterization of addressability by simultaneous randomized benchmarking},\ }\href {https://doi.org/https://doi.org/10.1103/PhysRevLett.109.240504} {\bibfield  {journal} {\bibinfo  {journal} {Physical review letters}\ }\textbf {\bibinfo {volume} {109}},\ \bibinfo {pages} {240504}
  (\bibinfo {year} {2012})}\BibitemShut {NoStop}%
\bibitem [{\citenamefont {Zakka-Bajjani}\ \emph {et~al.}(2011)\citenamefont {Zakka-Bajjani}, \citenamefont {Nguyen}, \citenamefont {Lee}, \citenamefont {Vale}, \citenamefont {Simmonds},\ and\ \citenamefont {Aumentado}}]{zakka2011quantum}%
  \BibitemOpen
  \bibfield  {author} {\bibinfo {author} {\bibfnamefont {E.}~\bibnamefont {Zakka-Bajjani}}, \bibinfo {author} {\bibfnamefont {F.}~\bibnamefont {Nguyen}}, \bibinfo {author} {\bibfnamefont {M.}~\bibnamefont {Lee}}, \bibinfo {author} {\bibfnamefont {L.~R.}\ \bibnamefont {Vale}}, \bibinfo {author} {\bibfnamefont {R.~W.}\ \bibnamefont {Simmonds}},\ and\ \bibinfo {author} {\bibfnamefont {J.}~\bibnamefont {Aumentado}},\ }\bibfield  {title} {\bibinfo {title} {Quantum superposition of a single microwave photon in two different’colour’states},\ }\href {https://doi.org/https://doi.org/10.1038/nphys2035} {\bibfield  {journal} {\bibinfo  {journal} {Nature Physics}\ }\textbf {\bibinfo {volume} {7}},\ \bibinfo {pages} {599} (\bibinfo {year} {2011})}\BibitemShut {NoStop}%
\bibitem [{\citenamefont {McKay}\ \emph {et~al.}(2017)\citenamefont {McKay}, \citenamefont {Wood}, \citenamefont {Sheldon}, \citenamefont {Chow},\ and\ \citenamefont {Gambetta}}]{mckay2017efficient}%
  \BibitemOpen
  \bibfield  {author} {\bibinfo {author} {\bibfnamefont {D.~C.}\ \bibnamefont {McKay}}, \bibinfo {author} {\bibfnamefont {C.~J.}\ \bibnamefont {Wood}}, \bibinfo {author} {\bibfnamefont {S.}~\bibnamefont {Sheldon}}, \bibinfo {author} {\bibfnamefont {J.~M.}\ \bibnamefont {Chow}},\ and\ \bibinfo {author} {\bibfnamefont {J.~M.}\ \bibnamefont {Gambetta}},\ }\bibfield  {title} {\bibinfo {title} {Efficient z gates for quantum computing},\ }\href {https://doi.org/https://doi.org/10.1103/PhysRevA.96.022330} {\bibfield  {journal} {\bibinfo  {journal} {Physical Review A}\ }\textbf {\bibinfo {volume} {96}},\ \bibinfo {pages} {022330} (\bibinfo {year} {2017})}\BibitemShut {NoStop}%
\bibitem [{\citenamefont {Gaebler}\ \emph {et~al.}(2012)\citenamefont {Gaebler}, \citenamefont {Meier}, \citenamefont {Tan}, \citenamefont {Bowler}, \citenamefont {Lin}, \citenamefont {Hanneke}, \citenamefont {Jost}, \citenamefont {Home}, \citenamefont {Knill}, \citenamefont {Leibfried} \emph {et~al.}}]{gaebler2012randomized}%
  \BibitemOpen
  \bibfield  {author} {\bibinfo {author} {\bibfnamefont {J.~P.}\ \bibnamefont {Gaebler}}, \bibinfo {author} {\bibfnamefont {A.~M.}\ \bibnamefont {Meier}}, \bibinfo {author} {\bibfnamefont {T.~R.}\ \bibnamefont {Tan}}, \bibinfo {author} {\bibfnamefont {R.}~\bibnamefont {Bowler}}, \bibinfo {author} {\bibfnamefont {Y.}~\bibnamefont {Lin}}, \bibinfo {author} {\bibfnamefont {D.}~\bibnamefont {Hanneke}}, \bibinfo {author} {\bibfnamefont {J.~D.}\ \bibnamefont {Jost}}, \bibinfo {author} {\bibfnamefont {J.}~\bibnamefont {Home}}, \bibinfo {author} {\bibfnamefont {E.}~\bibnamefont {Knill}}, \bibinfo {author} {\bibfnamefont {D.}~\bibnamefont {Leibfried}}, \emph {et~al.},\ }\bibfield  {title} {\bibinfo {title} {Randomized benchmarking of multiqubit gates},\ }\href {https://doi.org/https://doi.org/10.1103/PhysRevLett.108.260503} {\bibfield  {journal} {\bibinfo  {journal} {Physical review letters}\ }\textbf {\bibinfo {volume} {108}},\ \bibinfo {pages} {260503} (\bibinfo {year} {2012})}\BibitemShut {NoStop}%
\bibitem [{\citenamefont {C{\'o}rcoles}\ \emph {et~al.}(2013)\citenamefont {C{\'o}rcoles}, \citenamefont {Gambetta}, \citenamefont {Chow}, \citenamefont {Smolin}, \citenamefont {Ware}, \citenamefont {Strand}, \citenamefont {Plourde},\ and\ \citenamefont {Steffen}}]{corcoles2013process}%
  \BibitemOpen
  \bibfield  {author} {\bibinfo {author} {\bibfnamefont {A.~D.}\ \bibnamefont {C{\'o}rcoles}}, \bibinfo {author} {\bibfnamefont {J.~M.}\ \bibnamefont {Gambetta}}, \bibinfo {author} {\bibfnamefont {J.~M.}\ \bibnamefont {Chow}}, \bibinfo {author} {\bibfnamefont {J.~A.}\ \bibnamefont {Smolin}}, \bibinfo {author} {\bibfnamefont {M.}~\bibnamefont {Ware}}, \bibinfo {author} {\bibfnamefont {J.}~\bibnamefont {Strand}}, \bibinfo {author} {\bibfnamefont {B.~L.}\ \bibnamefont {Plourde}},\ and\ \bibinfo {author} {\bibfnamefont {M.}~\bibnamefont {Steffen}},\ }\bibfield  {title} {\bibinfo {title} {Process verification of two-qubit quantum gates by randomized benchmarking},\ }\href {https://doi.org/https://doi.org/10.1103/PhysRevA.87.030301} {\bibfield  {journal} {\bibinfo  {journal} {Physical Review A}\ }\textbf {\bibinfo {volume} {87}},\ \bibinfo {pages} {030301} (\bibinfo {year} {2013})}\BibitemShut {NoStop}%
\bibitem [{\citenamefont {Magesan}\ \emph {et~al.}(2012)\citenamefont {Magesan}, \citenamefont {Gambetta}, \citenamefont {Johnson}, \citenamefont {Ryan}, \citenamefont {Chow}, \citenamefont {Merkel}, \citenamefont {da~Silva}, \citenamefont {Keefe}, \citenamefont {Rothwell}, \citenamefont {Ohki}, \citenamefont {Ketchen},\ and\ \citenamefont {Steffen}}]{magesan2012interleavedrb}%
  \BibitemOpen
  \bibfield  {author} {\bibinfo {author} {\bibfnamefont {E.}~\bibnamefont {Magesan}}, \bibinfo {author} {\bibfnamefont {J.~M.}\ \bibnamefont {Gambetta}}, \bibinfo {author} {\bibfnamefont {B.~R.}\ \bibnamefont {Johnson}}, \bibinfo {author} {\bibfnamefont {C.~A.}\ \bibnamefont {Ryan}}, \bibinfo {author} {\bibfnamefont {J.~M.}\ \bibnamefont {Chow}}, \bibinfo {author} {\bibfnamefont {S.~T.}\ \bibnamefont {Merkel}}, \bibinfo {author} {\bibfnamefont {M.~P.}\ \bibnamefont {da~Silva}}, \bibinfo {author} {\bibfnamefont {G.~A.}\ \bibnamefont {Keefe}}, \bibinfo {author} {\bibfnamefont {M.~B.}\ \bibnamefont {Rothwell}}, \bibinfo {author} {\bibfnamefont {T.~A.}\ \bibnamefont {Ohki}}, \bibinfo {author} {\bibfnamefont {M.~B.}\ \bibnamefont {Ketchen}},\ and\ \bibinfo {author} {\bibfnamefont {M.}~\bibnamefont {Steffen}},\ }\bibfield  {title} {\bibinfo {title} {Efficient measurement of quantum gate error by interleaved randomized benchmarking},\ }\href {https://doi.org/10.1103/PhysRevLett.109.080505} {\bibfield  {journal}
  {\bibinfo  {journal} {Phys. Rev. Lett.}\ }\textbf {\bibinfo {volume} {109}},\ \bibinfo {pages} {080505} (\bibinfo {year} {2012})}\BibitemShut {NoStop}%
\bibitem [{\citenamefont {Abad}\ \emph {et~al.}(2022)\citenamefont {Abad}, \citenamefont {Fern{\'a}ndez-Pend{\'a}s}, \citenamefont {Frisk~Kockum},\ and\ \citenamefont {Johansson}}]{abad2022universal}%
  \BibitemOpen
  \bibfield  {author} {\bibinfo {author} {\bibfnamefont {T.}~\bibnamefont {Abad}}, \bibinfo {author} {\bibfnamefont {J.}~\bibnamefont {Fern{\'a}ndez-Pend{\'a}s}}, \bibinfo {author} {\bibfnamefont {A.}~\bibnamefont {Frisk~Kockum}},\ and\ \bibinfo {author} {\bibfnamefont {G.}~\bibnamefont {Johansson}},\ }\bibfield  {title} {\bibinfo {title} {Universal fidelity reduction of quantum operations from weak dissipation},\ }\href {https://doi.org/https://doi.org/10.1103/PhysRevLett.129.150504} {\bibfield  {journal} {\bibinfo  {journal} {Physical Review Letters}\ }\textbf {\bibinfo {volume} {129}},\ \bibinfo {pages} {150504} (\bibinfo {year} {2022})}\BibitemShut {NoStop}%
\bibitem [{\citenamefont {Kalfus}\ \emph {et~al.}(2020)\citenamefont {Kalfus}, \citenamefont {Lee}, \citenamefont {Ribeill}, \citenamefont {Fallek}, \citenamefont {Wagner}, \citenamefont {Donovan}, \citenamefont {Rist{\`e}},\ and\ \citenamefont {Ohki}}]{kalfus2020high}%
  \BibitemOpen
  \bibfield  {author} {\bibinfo {author} {\bibfnamefont {W.~D.}\ \bibnamefont {Kalfus}}, \bibinfo {author} {\bibfnamefont {D.~F.}\ \bibnamefont {Lee}}, \bibinfo {author} {\bibfnamefont {G.~J.}\ \bibnamefont {Ribeill}}, \bibinfo {author} {\bibfnamefont {S.~D.}\ \bibnamefont {Fallek}}, \bibinfo {author} {\bibfnamefont {A.}~\bibnamefont {Wagner}}, \bibinfo {author} {\bibfnamefont {B.}~\bibnamefont {Donovan}}, \bibinfo {author} {\bibfnamefont {D.}~\bibnamefont {Rist{\`e}}},\ and\ \bibinfo {author} {\bibfnamefont {T.~A.}\ \bibnamefont {Ohki}},\ }\bibfield  {title} {\bibinfo {title} {High-fidelity control of superconducting qubits using direct microwave synthesis in higher nyquist zones},\ }\href {https://doi.org/10.1109/TQE.2020.3042895} {\bibfield  {journal} {\bibinfo  {journal} {IEEE Transactions on Quantum Engineering}\ }\textbf {\bibinfo {volume} {1}},\ \bibinfo {pages} {1} (\bibinfo {year} {2020})}\BibitemShut {NoStop}%
\bibitem [{\citenamefont {Ryan}\ \emph {et~al.}(2017)\citenamefont {Ryan}, \citenamefont {Johnson}, \citenamefont {Rist{\`e}}, \citenamefont {Donovan},\ and\ \citenamefont {Ohki}}]{ryan2017hardware}%
  \BibitemOpen
  \bibfield  {author} {\bibinfo {author} {\bibfnamefont {C.~A.}\ \bibnamefont {Ryan}}, \bibinfo {author} {\bibfnamefont {B.~R.}\ \bibnamefont {Johnson}}, \bibinfo {author} {\bibfnamefont {D.}~\bibnamefont {Rist{\`e}}}, \bibinfo {author} {\bibfnamefont {B.}~\bibnamefont {Donovan}},\ and\ \bibinfo {author} {\bibfnamefont {T.~A.}\ \bibnamefont {Ohki}},\ }\bibfield  {title} {\bibinfo {title} {Hardware for dynamic quantum computing},\ }\bibfield  {journal} {\bibinfo  {journal} {Review of Scientific Instruments}\ }\textbf {\bibinfo {volume} {88}},\ \href {https://doi.org/https://doi.org/10.1063/1.5006525} {https://doi.org/10.1063/1.5006525} (\bibinfo {year} {2017})\BibitemShut {NoStop}%
\bibitem [{\citenamefont {You}\ \emph {et~al.}(2019)\citenamefont {You}, \citenamefont {Sauls},\ and\ \citenamefont {Koch}}]{you2019circuit}%
  \BibitemOpen
  \bibfield  {author} {\bibinfo {author} {\bibfnamefont {X.}~\bibnamefont {You}}, \bibinfo {author} {\bibfnamefont {J.~A.}\ \bibnamefont {Sauls}},\ and\ \bibinfo {author} {\bibfnamefont {J.}~\bibnamefont {Koch}},\ }\bibfield  {title} {\bibinfo {title} {Circuit quantization in the presence of time-dependent external flux},\ }\href {https://doi.org/https://doi.org/10.1103/PhysRevB.99.174512} {\bibfield  {journal} {\bibinfo  {journal} {Physical Review B}\ }\textbf {\bibinfo {volume} {99}},\ \bibinfo {pages} {174512} (\bibinfo {year} {2019})}\BibitemShut {NoStop}%
\bibitem [{\citenamefont {Bryon}\ \emph {et~al.}(2023)\citenamefont {Bryon}, \citenamefont {Weiss}, \citenamefont {You}, \citenamefont {Sussman}, \citenamefont {Croot}, \citenamefont {Huang}, \citenamefont {Koch},\ and\ \citenamefont {Houck}}]{bryon2023time}%
  \BibitemOpen
  \bibfield  {author} {\bibinfo {author} {\bibfnamefont {J.}~\bibnamefont {Bryon}}, \bibinfo {author} {\bibfnamefont {D.}~\bibnamefont {Weiss}}, \bibinfo {author} {\bibfnamefont {X.}~\bibnamefont {You}}, \bibinfo {author} {\bibfnamefont {S.}~\bibnamefont {Sussman}}, \bibinfo {author} {\bibfnamefont {X.}~\bibnamefont {Croot}}, \bibinfo {author} {\bibfnamefont {Z.}~\bibnamefont {Huang}}, \bibinfo {author} {\bibfnamefont {J.}~\bibnamefont {Koch}},\ and\ \bibinfo {author} {\bibfnamefont {A.~A.}\ \bibnamefont {Houck}},\ }\bibfield  {title} {\bibinfo {title} {Time-dependent magnetic flux in devices for circuit quantum electrodynamics},\ }\href {https://doi.org/https://doi.org/10.1103/PhysRevApplied.19.034031} {\bibfield  {journal} {\bibinfo  {journal} {Physical Review Applied}\ }\textbf {\bibinfo {volume} {19}},\ \bibinfo {pages} {034031} (\bibinfo {year} {2023})}\BibitemShut {NoStop}%
\bibitem [{\citenamefont {Riwar}\ and\ \citenamefont {DiVincenzo}(2022)}]{riwar2022circuit}%
  \BibitemOpen
  \bibfield  {author} {\bibinfo {author} {\bibfnamefont {R.-P.}\ \bibnamefont {Riwar}}\ and\ \bibinfo {author} {\bibfnamefont {D.~P.}\ \bibnamefont {DiVincenzo}},\ }\bibfield  {title} {\bibinfo {title} {Circuit quantization with time-dependent magnetic fields for realistic geometries},\ }\href {https://doi.org/https://doi.org/10.1038/s41534-022-00539-x} {\bibfield  {journal} {\bibinfo  {journal} {npj Quantum Information}\ }\textbf {\bibinfo {volume} {8}},\ \bibinfo {pages} {36} (\bibinfo {year} {2022})}\BibitemShut {NoStop}%
\bibitem [{\citenamefont {Home}\ \emph {et~al.}(2009)\citenamefont {Home}, \citenamefont {Hanneke}, \citenamefont {Jost}, \citenamefont {Amini}, \citenamefont {Leibfried},\ and\ \citenamefont {Wineland}}]{home2009complete}%
  \BibitemOpen
  \bibfield  {author} {\bibinfo {author} {\bibfnamefont {J.~P.}\ \bibnamefont {Home}}, \bibinfo {author} {\bibfnamefont {D.}~\bibnamefont {Hanneke}}, \bibinfo {author} {\bibfnamefont {J.~D.}\ \bibnamefont {Jost}}, \bibinfo {author} {\bibfnamefont {J.~M.}\ \bibnamefont {Amini}}, \bibinfo {author} {\bibfnamefont {D.}~\bibnamefont {Leibfried}},\ and\ \bibinfo {author} {\bibfnamefont {D.~J.}\ \bibnamefont {Wineland}},\ }\bibfield  {title} {\bibinfo {title} {Complete methods set for scalable ion trap quantum information processing},\ }\href {https://doi.org/https://doi.org/10.1126/science.1177077} {\bibfield  {journal} {\bibinfo  {journal} {Science}\ }\textbf {\bibinfo {volume} {325}},\ \bibinfo {pages} {1227} (\bibinfo {year} {2009})}\BibitemShut {NoStop}%
\bibitem [{\citenamefont {Brown}\ \emph {et~al.}(2022)\citenamefont {Brown}, \citenamefont {Doucet}, \citenamefont {Rist{\`e}}, \citenamefont {Ribeill}, \citenamefont {Cicak}, \citenamefont {Aumentado}, \citenamefont {Simmonds}, \citenamefont {Govia}, \citenamefont {Kamal},\ and\ \citenamefont {Ranzani}}]{brown2022trade}%
  \BibitemOpen
  \bibfield  {author} {\bibinfo {author} {\bibfnamefont {T.}~\bibnamefont {Brown}}, \bibinfo {author} {\bibfnamefont {E.}~\bibnamefont {Doucet}}, \bibinfo {author} {\bibfnamefont {D.}~\bibnamefont {Rist{\`e}}}, \bibinfo {author} {\bibfnamefont {G.}~\bibnamefont {Ribeill}}, \bibinfo {author} {\bibfnamefont {K.}~\bibnamefont {Cicak}}, \bibinfo {author} {\bibfnamefont {J.}~\bibnamefont {Aumentado}}, \bibinfo {author} {\bibfnamefont {R.}~\bibnamefont {Simmonds}}, \bibinfo {author} {\bibfnamefont {L.}~\bibnamefont {Govia}}, \bibinfo {author} {\bibfnamefont {A.}~\bibnamefont {Kamal}},\ and\ \bibinfo {author} {\bibfnamefont {L.}~\bibnamefont {Ranzani}},\ }\bibfield  {title} {\bibinfo {title} {Trade off-free entanglement stabilization in a superconducting qutrit-qubit system},\ }\href {https://doi.org/https://doi.org/10.1038/s41467-022-31638-0} {\bibfield  {journal} {\bibinfo  {journal} {Nature communications}\ }\textbf {\bibinfo {volume} {13}},\ \bibinfo {pages} {3994} (\bibinfo {year} {2022})}\BibitemShut {NoStop}%
\bibitem [{\citenamefont {Sete}\ \emph {et~al.}(2021)\citenamefont {Sete}, \citenamefont {Didier}, \citenamefont {Chen}, \citenamefont {Kulshreshtha}, \citenamefont {Manenti},\ and\ \citenamefont {Poletto}}]{sete2021parametric}%
  \BibitemOpen
  \bibfield  {author} {\bibinfo {author} {\bibfnamefont {E.~A.}\ \bibnamefont {Sete}}, \bibinfo {author} {\bibfnamefont {N.}~\bibnamefont {Didier}}, \bibinfo {author} {\bibfnamefont {A.~Q.}\ \bibnamefont {Chen}}, \bibinfo {author} {\bibfnamefont {S.}~\bibnamefont {Kulshreshtha}}, \bibinfo {author} {\bibfnamefont {R.}~\bibnamefont {Manenti}},\ and\ \bibinfo {author} {\bibfnamefont {S.}~\bibnamefont {Poletto}},\ }\bibfield  {title} {\bibinfo {title} {Parametric-resonance entangling gates with a tunable coupler},\ }\href {https://doi.org/https://doi.org/10.1103/PhysRevApplied.16.024050} {\bibfield  {journal} {\bibinfo  {journal} {Physical Review Applied}\ }\textbf {\bibinfo {volume} {16}},\ \bibinfo {pages} {024050} (\bibinfo {year} {2021})}\BibitemShut {NoStop}%
\end{thebibliography}%

\end{document}